\documentclass[fleqn,usenatbib]{mnras}

\usepackage{amsmath}
\usepackage{booktabs}
\usepackage{enumitem}
\usepackage{tablefootnote}
\usepackage{comment}

\usepackage{tikz}
\usepackage{tikz-3dplot}

\tdplotsetmaincoords{60}{110}
\pgfmathsetmacro{\rvec}{.8}
\pgfmathsetmacro{\thetavec}{30}
\pgfmathsetmacro{\phivec}{60}

\DeclareRobustCommand{\VAN}[3]{#2}
\let\VANthebibliography\thebibliography
\def\thebibliography{\DeclareRobustCommand{\VAN}[3]{##3}\VANthebibliography}

\usepackage{graphicx}	% Including figure files
\usepackage{url}
\usepackage{multirow}

\usepackage{newtxtext,newtxmath}

\title[The Zeeman Effect in Protostellar Envelopes]{Modeling CN Zeeman Effect Observations of the Envelopes of a Low-Mass Protostellar Disk and a Massive Protostar}

\author[R. Mazzei et al.]{
Renato Mazzei,$^{1}$\thanks{E-mail: mazzei@virginia.edu}
Zhi-Yun Li,$^{1}$
Che-Yu Chen,$^{1,2}$
Yisheng Tu,$^{1}$ 
Laura Fissel,$^{3}$
and Richard I. Klein$^{2,4}$
\\
$^{1}$Department of Astronomy, University of Virginia, 530 McCormick Rd., Charlottesville, Virginia 22904, USA\\
$^{2}$Lawrence Livermore National Laboratory, Livermore, CA 94550, USA\\
$^{3}$Department of Physics, Engineering Physics, and Astronomy, Queen's University, Kingston, Ontario, Canada\\
$^{4}$Department of Astronomy, University of California, Berkeley, CA 94720, USA
}

\date{Accepted XXX. Received YYY; in original form ZZZ}

\pubyear{2023}

\begin{document}
\label{firstpage}
\pagerange{\pageref{firstpage}--\pageref{lastpage}}
\maketitle

% Abstract of the paper
\begin{abstract}
We use the POLARIS radiative transfer code to produce simulated circular polarization Zeeman emission maps of the CN $J = 1 - 0$ molecular line transition for two types of protostellar envelope magnetohydrodynamic simulations.
Our first model is a low mass disk envelope system (box length $L = 200\text{ au}$), and our second model is the envelope of a massive protostar ($L = 10^4\text{ au}$) with a protostellar wind and a CN enhanced outflow shell.
We compute the velocity-integrated Stokes $I$ and $V$, as well as the implied $V/I$ polarization percentage, for each detector pixel location in our simulated emission maps.
Our results show that both types of protostellar environment are in principle accessible with current circular polarization instruments, with each containing swaths of envelope area that yield percentage polarizations that exceed the 1.8\% nominal sensitivity limit for circular polarization experiments with the Atacama Large Millimeter/submillimeter Array (ALMA).
In both systems, high polarization ($\gtrsim$1.8\%) pixels tend to lie at an intermediate distance away from the central star and where the line-center opacity of the CN emission is moderately optically thin ($\tau_{LC} \sim 0.1-1$).
Furthermore, our computed $V/I$ values scale roughly with the density weighted mean line-of-sight magnetic field strength, indicating that Zeeman observations can effectively diagnose the strength of envelope-scale magnetic fields.  
We also find that pixels with large $V/I$ are preferentially co-located where the absolute value of the velocity-integrated $V$ is also large, suggesting that locations with favorable percentage polarization are also favorable in terms of raw signal.

\end{abstract}

% Select between one and six entries from the list of approved keywords.
% Don't make up new ones.

%%%%%%%%%%%%%%%%%%%%%%%%%%%%%%%%%%%%%%%%%%%%%%%%%%

%%%%%%%%%%%%%%%%% BODY OF PAPER %%%%%%%%%%%%%%%%%%

\section{Introduction}

Magnetic fields are pervasive across many scales in star formation environments and are expected to play an important role in regulating gas flows and structure formation at a variety of stages of the star formation process.
On the molecular cloud scale ($\gtrsim$1 pc), magnetic fields both restrict gas flows and provide opposition to gravitational collapse \citep{mestel1956,mouschovias1976}.   
Cloud-scale fields are well studied with far-infrared/sub-mm polarimetry experiments such as Planck, BLASTPol, and SOFIA \citep{planckXIX,galitzki2014,fissel2016,chuss2019}.
These experiments leverage the well-established ``radiative torques" (RAT) grain alignment theory, wherein spinning, effectively oblate dust grains preferentially align with their short axes parallel to the local magnetic field direction \citep{lazarian2007}, yielding linear polarization that is perpendicular to the magnetic field \citep{davis1951}.

While dust grain alignment with the magnetic field is understood to be the dominant source of linearly polarized far-IR emission at larger scales, 
at smaller scales ($\sim$100s-1000 au), however, the interpretation of linear polarization as a magnetic field tracer is more tenuous because other sources of polarization become important.
In disks particularly, self-scattering of thermal dust emission \citep{kataoka2015,yang2016}, gas flow alignment \citep{kataoka2019}, and ``k-RAT" radiation field alignment \citep{kataoka2017, tazaki2017} may all produce linear polarization signatures.
Nonetheless, observing magnetic fields on these smaller scales remains of great interest.
In older sources (Class II), magnetic fields are predicted to launch jets and winds along the disk axis \citep{blandford1982}, generate magneto-rotational instability \citep[MRI;][]{balbus1991}, and produce flows that contribute to the formation of rings and gaps \citep{suriano2017} or planetesimals \citep[via zonal flows;][]{johansen2009}.
In younger (Class 0/I) protostars, magnetic fields may also hinder disk growth on 100+ au scales through magnetic braking, especially when the magnetic field aligns with the rotation axis of the surrounding envelope \citep{mellon2008, hennebelle2008}.

Another method for accessing magnetic fields in small-scale sources is to observe molecular line transitions in species that are sensitive to the Zeeman Effect (e.g., CN, OH, HI).
In the presence of a magnetic field, the energy levels of these lines are split into higher and lower energy circularly polarized components, and the degree of splitting is proportional to the line-of-sight magnetic field strength \citep{crutcher1993}.
Historically, Zeeman observations have mainly been performed using single-dished telescopes to probe core-scale ($\sim$0.1 pc) line-of-sight magnetic field strengths \citep[see, e.g.,][]{heiles2004,falgarone2008,troland2008,crutcher2010}.
However, in recent years with the advent of a circular polarization observing mode on the Atacama Large Millimeter/submillimeter Array (ALMA), higher resolution ($\lesssim$1$^{\prime \prime}$) Zeeman observations are now possible.
Disk-scale observations have proven challenging, with programs to-date yielding upper limit constraints only \citep{vlemmings2019,harrison2021}.
Though disk-scale magnetic field strengths of up to $\sim$3 mG or more are predicted, the line-of-sight strength is likely suppressed in many disks due to cancellation within the torodial $\boldsymbol{B}$-field component \citep{mazzei2020,lankhaar2023}.

Given the complicated situation inside disks, an alternative approach for searching for magnetic field signatures in protostellar environments is to probe the inner envelope regions beyond the edge of the disk.
The advantages of the envelope are two-fold.
First, the inner envelope-scale magnetic field is expected to be almost as strong as in a disk, while still remaining relatively unaffected by disk-scale tangling and toroidal wind-up.
Linear dust polarization studies of these environments have revealed polarization percentages up to $\sim$5$\%$, with relatively uniform geometry in some cases \citep{cox2018}.
Second, some envelope sources have been observed to have bright emission of Zeeman-sensitive molecules such as CN \citep[see, e.g.,][]{tychoniec2021}.

In this work, we conduct 3D radiative transfer simulations to produce simulated emission maps of circularly polarized emission of the CN $J = 1-0$ transition for two different simulated protostellar envelope sources.
Our first test case is a protostellar disk around a low-mass protostar, and the second the envelope around a massive protostar. 
For each simulation, we calculate the Stokes $I$ and Stokes $V$ obtained from many beams 
across the envelopes of our sources.
We report the percentage polarization ($V/I$) and compare the implied line-of-sight magnetic field strength ($B_{\rm LOS}$) from the Zeeman fitting with the actual mean (density-weighted) $B_{\rm LOS}$.
There is substantial variability in the results for different beams due to non-uniform local magnetic field structure, but in some cases we find percentage polarization values $\gtrsim$2\%. 

Given the nominal ALMA sensitivity limit of 1.8\%, this suggests that Zeeman experiments with current instruments can be a useful way to study magnetic fields in (proto)stellar envelopes.
However, program success depends critically on source selection and beam placement.
It should be noted that our comparisons in this work are based strictly on radiative transfer simulations; we do not use the ALMA simulator to produce simulated observable products.
Direct observational comparison of simulations to ALMA program data should involve use of the \textsc{CASA} simulator \citep{casa2007}.
This is outside the scope of our work here.
Nonetheless, from our results we are able to identify some general criteria for setting up observations to maximize detectability.
Particularly, we find that regions some distance away from the central star at intermediate line-center optical depth ($\tau
_{LC} \sim 0.1 -1$) tend to be favorable.
Furthermore, our results suggest that continued improvement of circular polarization instruments will be extremely fruitful; we predict that a factor of ten improvement in sensitivity (i.e., to a 0.18\% limit) will make sightlines across nearly the entire inner envelope detectable for sources with typical magnetic field strengths (i.e., $\gtrsim 1 \text{ mG}$, comparable to those found in our simulations).

This paper is organized as follows.
In Section \ref{sec:sims} we introduce the two model types we consider in this work, then in Section \ref{sec:zeeman} we discuss our methods for making our simulated Zeeman emission maps.
Section \ref{sec:results} presents the results of our simulations, including maps of the observables (Sec.~\ref{ssec:results_maps}), computations of percentage polarization values through the envelopes of our sources (Sec.~\ref{ssec:results_pp}), and comparisons of our pixel-by-pixel percentage polarizations to other associated observables as well as the underlying magnetic field information from our MHD simulations (Sec.~\ref{ssec:results_2dhist}).
We then provide some discussion in Section \ref{sec:disc} on some additional criteria that can affect whether a given sightline or location is favorable for Zeeman experiments, including some auxiliary observational factors of which observers should also be aware (Sec.~\ref{ssec:disc_obscav}). 
Finally, the main conclusions of this work are summarized in Section \ref{sec:sum}.

\section{Numerical Simulations} \label{sec:sims}

Both sets of simulations studied in this work were performed using 3D grid-based MHD codes.

\subsection{Low-Mass Protostellar Disk Envelope Simulation} \label{ssec:stenv}

Our first model type is an envelope and disk system around a low-mass protostar (henceforth known as \texttt{lmde}).
The turbulent, non-ideal MHD model we used was the reference model of \citet{tu2023arxiv}.
Here we highlight a few salient features of the model and refer the reader to \citet{tu2023arxiv} for details.
We use the ATHENA++ code \citep{stone2008,stone2020}, and include adaptive mesh refinement (AMR) and full multigrid (FMG) self-gravity \citep{tomidastone2023}.

The physical setup is similar to that used in \citet{lam2019}. 
We initialize a pseudo-Bonner-Ebert sphere prescribed by

\begin{equation}
    \rho(r) = \frac{\rho_0}{1+(r/r_c)^2}\,.
    \label{eq:bonnerSphere}
\end{equation}
%In our ``fiducial" model (henceforth known as \texttt{lmde}), 
We set the central density $\rho_0 = 4.6 \times 10^{-17}$ g cm$^{-3}$, and characteristic radius $r_c = 670$ au.
The sphere is embedded in a diffuse low-density gas ($\rho_{\rm amb} = 4.56\times10^{-20}$ g cm$^{-3}$).
The total box size of the simulation domain is 10,000 au per side length, and the total gas mass is 0.56 $M_{\odot}$.
The simulation is initialized with $k^{-2}$ power spectrum turbulence \citep{kolmogorov1941, gong2011} with an rms Mach number of unity, solid body rotation rate $\boldsymbol{\omega} = (\omega_x,\omega_y,\omega_z) = (0,0,6.16\times10^{-13}\text{ s}^{-1})$, and uniform magnetic field $\boldsymbol{B} = (B_x,B_y,B_z) = (0,0,2.2\times10^{-4}\text{ G})$. 
The rotation rate corresponds to a ratio of rotational to gravitational energies of 0.03, which is in the range of the values inferred observationally by \citet{goodman1993} and \citet{caselli2002} through velocity gradients across dense cores. However, what fraction of the measured gradient is contributed by rotation remains uncertain. The adopted field strength corresponds to a dimensionless mass-to-flux ratio of 2.6, close to the median value for dense cores inferred by \citet{troland2008} through OH Zeeman measurements after geometric corrections. We note that although the magnetic field is initially uniform, it is significantly distorted by the imposed turbulence, gravitational collapse, and rotation at later times, particularly in the inner envelope surrounding the disk [see, e.g., Fig.~4a of \citet{tu2023arxiv}].

Ambipolar diffusivity $\eta_A$ is prescribed as
\begin{equation}
    \eta_A = \eta_0 \frac{B^2}{4\pi \rho^{3/2}}\,.
    \label{eq:ambipolar}
\end{equation}
Based on cosmic ionization rate calculations from \citet{shu1992}, we set $\eta_0 = 95.2\text{ g}^{1/2}\text{ cm}^{-3/2}$, corresponding to a standard reference value.

\begin{figure*}
\centering
\includegraphics[width=\textwidth]{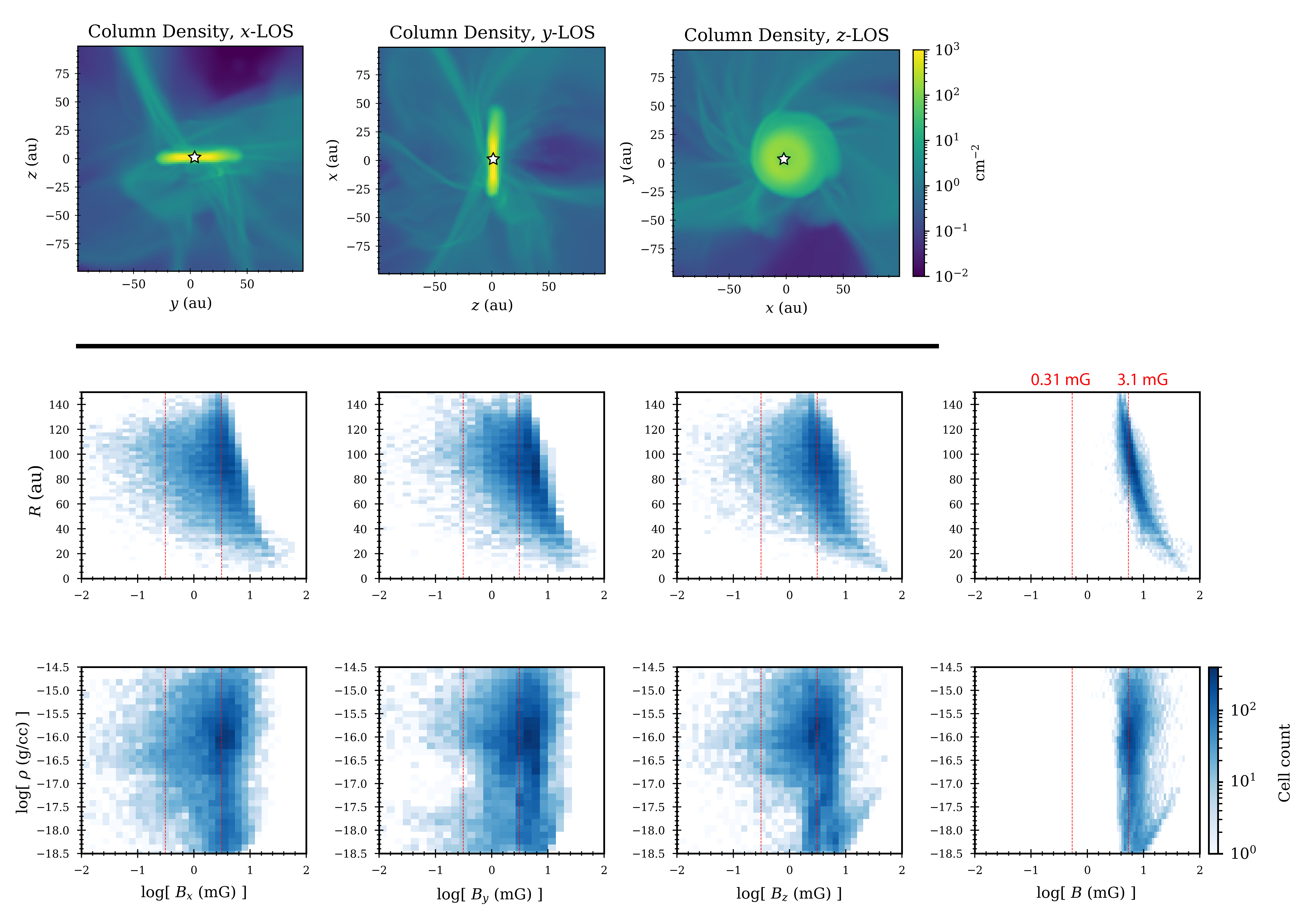}
\caption{\textit{\textbf{Top row:}} Column density plots for each of the Cartesian axis views of our \texttt{lmde} model. \textit{\textbf{Middle row:}} 2-dimensional histogram plots of local magnetic field component strengths (in the final column, $B = \sqrt{B_x^2 + B_y^2 + B_z^2}$) versus distance from the central sink particle. Annotated in red are lines corresponding to 3.1 mG and 0.31 mG line-of-sight magnetic field strength  (Section \ref{sec:zeeman} discusses why these values are highlighted). \textit{\textbf{Bottom row:}} 2-dimensional histogram plots of local magnetic field component strengths versus local density.} 
\label{fig:y2dhist}
\end{figure*}

We perform our polarization analysis on a simulation frame at $t = 3 \times 10^4$ yr.
By this time in the simulation, a stable disk has formed.
The disk remains stable for a while after this as well, but slowly decreases in mass at later times \citep{tu2023arxiv}. Our chosen frame corresponds roughly to the time of maximum disk mass. 
Column density plots of the \texttt{lmde} model for each the $x-$, $y-$, and $z-$lines-of-sight are shown in Figure \ref{fig:y2dhist}, along with 2-dimensional histograms of magnetic field strength versus gas density and distance from the (central) sink particle\footnote{We note that the field strength in the inner protostellar envelope depends in a complex way on several factors, including the initial core mass and field strength, level of turbulence, and magnetic diffusivities. For example, \citet{mignonrisse2021} simulated magnetized disk formation in more massive cores with turbulence. They found milli-Gauss (mG) magnetic fields up to 1000 au, with the field strength reaching 100 mG or more on the inner ~100 au scale (see their Fig.~13). Their magnetic field is stronger than ours, making it more detectable in principle. We postpone a detailed exploration of parameter space to a future investigation.}.

\begin{figure}
\centering
\includegraphics[width=\columnwidth]{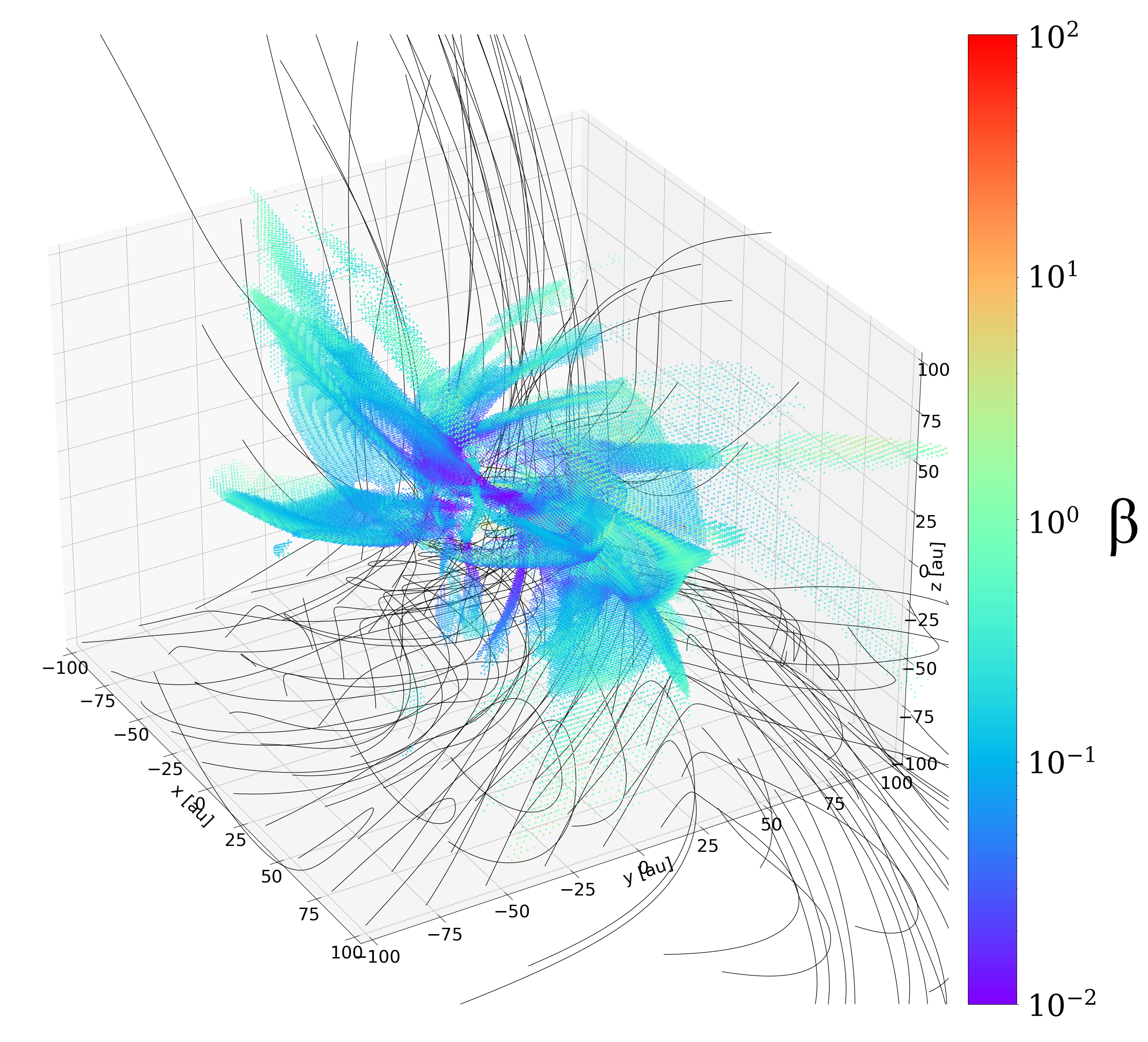}
\caption{3-dimensional view of magnetic field lines in the inner protostellar envelope region of our \texttt{lmde} model. Also included are 3D contours of the dimensionless plasma-$\beta$ to highlight the flattened dense demagnetized structures that dominate the dynamics of the inner envelope [see \citet{tu2023arxiv} for detailed discussion].} 
\label{fig:y3dstream}
\end{figure}

In Figure \ref{fig:y3dstream}, we provide a 3-dimensional view of the magnetic field lines in the envelope. Whereas the magnetic field inside the disk is dominated by the toroidal component because of rotational wrapping [not shown here, but see discussion in \citet{tu2023arxiv}], there are regions in the envelope where the field lines are more uniform and well-behaved.

\subsection{Massive Star-forming Envelope (MSEnv) Simulation} \label{ssec:msenv}

The massive star formation simulation (henceforth referred to as model \texttt{MSEnv}) considered here was conducted using the Radiation-Magnetohydrodynamic (RMHD) code \textsc{Orion2} \citep{ORION2} with 5 levels of AMR (highest resolution $\Delta x \approx 18$\,au). 
The setup is similar to those described in \cite{AJC_2011} but now with magnetic fields.
The simulation was initialized as a massive cloud core of mass $M_{\rm core} = 10^3$\,M$_\odot$ with a power-law density profile $\rho(r) \propto r^{-3/2}$ up to the radius of the core $R_{\rm core} = 0.18$\,pc, which gives an average column density of $2.0$\,g$\,$cm$^{-2}$. 
We then add a turbulent velocity field to the core with rms Mach number ${\cal M} = 2.43$, which corresponds to a virial parameter $\alpha_{\rm vir} = 1.67$ for the cloud core. 
This value was chosen for the balance between gravity and turbulent motions during the protostellar system evolution; see \cite{McKeeTan03} for more detailed discussions.
The initial magnetic field strength of the core is chosen so that the dimensionless mass-to-flux ratio is $3.0$ inside the core, slightly larger than those observed in high-mass star-forming clumps \citep[$\sim 1.5-2$; see e.g.,][]{crutcher2012, Pillai_CNZeeman_16, Motte_HMSFreview_18} to ensure gravitational collapse.  
The initial gas temperature inside the core is $T_{\rm gas} = 35$\,K, which is also set to be the temperature floor of the simulation box to avoid numerical rarefaction. 
A frequency-integrated flux limited diffusion (FLD) algorithm is adopted to approximate the radiation transport (see \citealt{AJC_2011} for more details).

In \textsc{Orion2}, the formation and evolution of protostars are handled by the star particle model described in \cite{Offner_2009}, which includes launching protostellar outflows. A tracer particle routine is implemented to record the properties of the ejected gas \citep{Offner_2009, AJC_2011}, and we utilized this function to trace the effective region of protostellar outflows in our simulated emission maps (see Sec.~\ref{sec:zeeman}).

\begin{figure*}
\centering
\includegraphics[width=\textwidth]{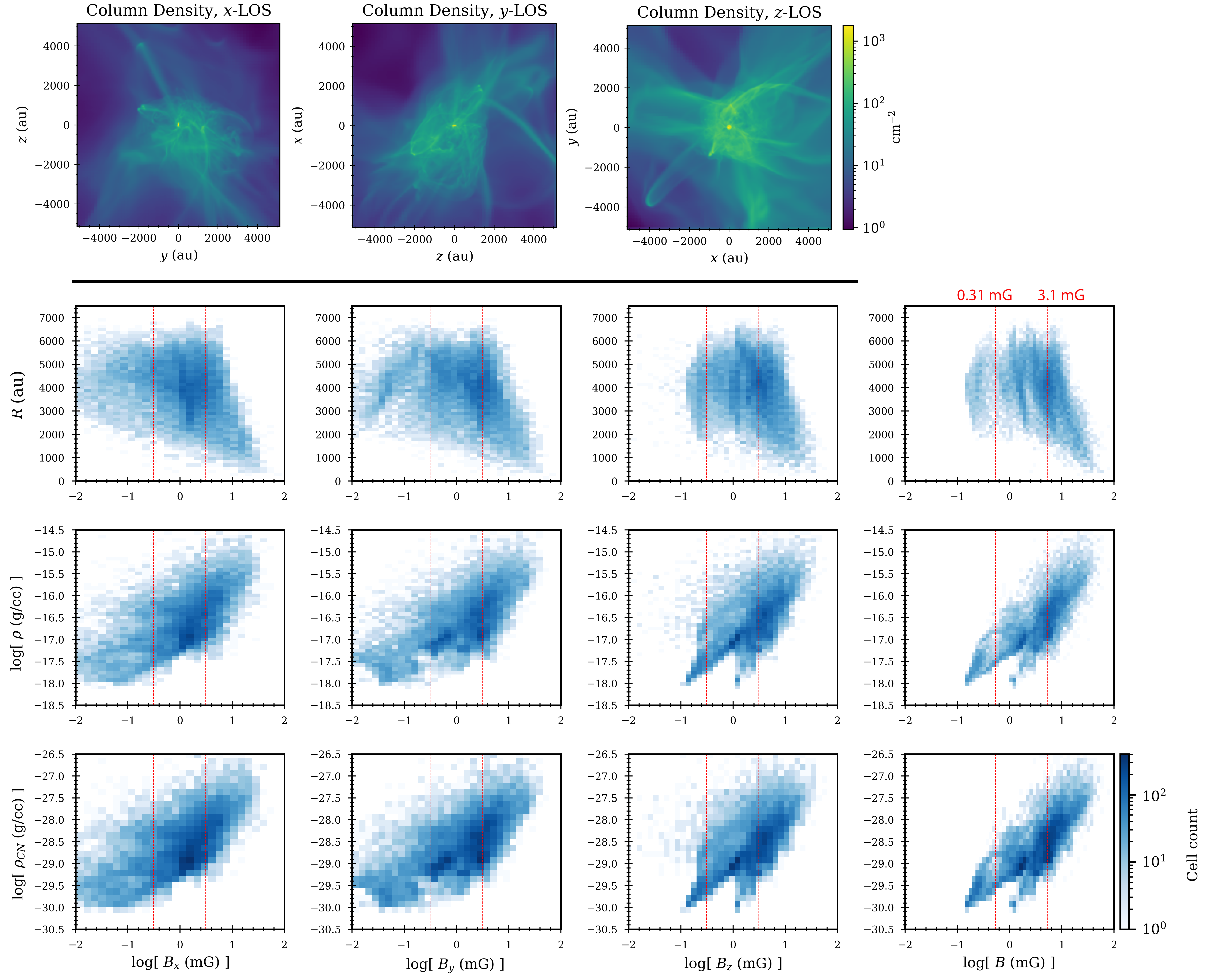}
\caption{\textit{\textbf{Top row:}} Column density plots for each of the Cartesian axis views of our \texttt{MSEnv} model. \textit{\textbf{2nd row:}} 2-dimensional histogram plots of local magnetic field component strengths (in the final column, $B = \sqrt{B_x^2 + B_y^2 + B_z^2}$) versus distance from the central sink particle. Annotated in red are lines corresponding to 3.1 mG and 0.31 mG line-of-sight magnetic field strength (Section \ref{sec:zeeman} discusses why these values are highlighted). \textit{\textbf{3rd row:}} 2-dimensional histogram plots of local magnetic field component strengths versus local density. \textit{\textbf{Bottom row:}} 2-dimensional histogram plots of local magnetic field component versus density, only including cells that are given a factor of 1000 enhancement in CN abundance in our model (from $X_{\rm amb} = 10^{-12}$ to $X_{\rm shell} = 10^{-9}$).} 
\label{fig:c2dhist}
\end{figure*}

We investigate the frame when the central star is about $29.8$\,M$_\odot$ in mass, focusing on a $L_{\rm box} = 10^{4}$\,au region around the massive protostar.
For easier computation we reproject the AMR data to $256^3$ unigrid array in \texttt{python} using \texttt{yt} \citep{yt2011}. 
In Figure \ref{fig:c2dhist} we present column density plots and 2-dimensional histograms of magnetic field and density data for model \texttt{MSEnv}.

\section{Simulated Zeeman Emission Maps} \label{sec:zeeman}
To characterize the observable magnetic field strength in each of the physical environments discussed above in Section \ref{sec:sims}, we produce simulated circular polarization emission maps of the CN $J = 1 - 0$ molecular line at 113 GHz.
This line comprises a suite of nine hyperfine components\footnote{See \citet{falgarone2008} for a full list of these components, including their rest frequencies, relative intensities, and Zeeman factor values.} that are non-blended and stackable \citep{mazzei2020}.
For the majority of this work we choose to only simulate one representative sub-transition, the 113.144 GHz component with relative intensity $\text{RI} = 8$ and Zeeman factor $z_B = 2.18$.
In Section \ref{ssec:disc_obscav_st} we consider the potential signal boost that can be gained from stacking.

We perform Zeeman-splitting line emission calculations using the POLARIS raditaive transfer code with the ZRAD extension \citep{brauer2017b}, which incorporates data from the Leiden Atomic and Molecular DAtabase \citep[LAMDA;][]{schoier2005} and the JPL spectral line catalog \citep{pickett1998}.
A Faddeeva function solver\footnote{\url{http://ab-initio.mit.edu/wiki/index.php/Faddeeva\_Package}, Copyright \copyright 2012 Massachusetts Institute of Technology} is used to compute the final line shape, and included in the calculations are considerations for natural, collisional, and Doppler broading, as well as the magneto-optic effect \citep{larsson2014}.

\begin{figure*}
\centering
\includegraphics[width=\textwidth]{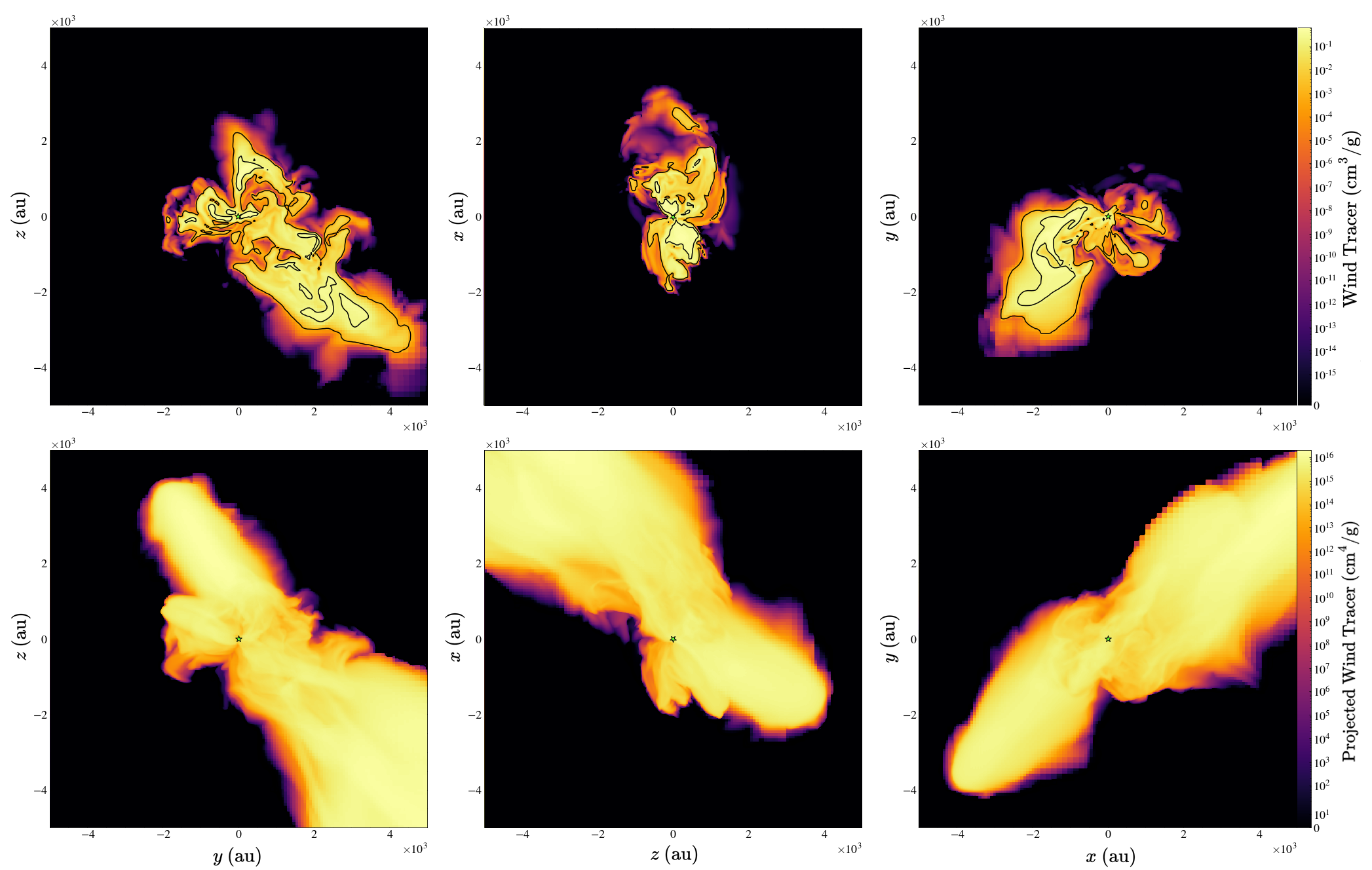}
\caption{\textit{\textbf{Top row:}} Midplane slice plots of the wind tracer value for the \texttt{MSEnv} model, as viewed from the $z-$, $y-$, and $x-$ lines-of-sight (left-to-right). Black contours are placed at wind tracer values of 0.001 and 0.2 cm$^{3}$ g$^{-1}$, the limits that define our CN enhanced shell.
\textit{\textbf{Bottom row:}} Projection plots of the line-of-sight integrated wind tracer value. 
}
\label{fig:shell}
\end{figure*}

Our POLARIS line radiative transfer calculations are computed on $128^3$ and $256^3$ fixed resolution grids for models \texttt{lmde} and \texttt{MSEnv}.
These translate to resolutions of 1.56 au and 39.1 au, respectively.
For each cell, local gas density, gas velocity components, and magnetic field components are supplied from our MHD simulations.
We also must specify a CN abundance value for each cell.
For our \texttt{lmde} model we choose constant CN abundance $X_{\rm CN} = 10^{-9}$.
This value ensures a moderately optically thin envelope beyond the disk edge at $R\sim$30 au, and is similar to estimates for CN abundances in disk modeling \citep{cazzoletti2018}.
In the massive star (\texttt{MSEnv}) model, we also set $X_{\rm CN} = 10^{-9}$, but only within a ``shell'' region that has cells containing tracer particles with values in a given range.
This is meant to simulate a CN enhancement at the edge of a cavity blown out by a protostellar wind that is more directly exposed to the UV radiation from the central massive protostar system. Such regions are observed in some cases to be enhanced due to far-ultraviolet dissociation chemistry \citep{arulanantham2020}. 
We set the range of tracer particle that define the shell by hand, choosing values that empirically produce a reasonable result.
Slice and projection plots of our shell are shown in Figure \ref{fig:shell}. 
For the remainder of the massive star envelope (i.e., all cells not within the shell), we set a lower ambient CN abundance $X_{\rm CN, amb} = 10^{-12}$.

Each line emission simulation converts the position-position-position (PPP) data cube into a position-position-velocity (PPV) data cube.
Values for the Stokes $I$, Stokes $V$, and optical depth $\tau$ are recorded in each pixel for 169 velocity channels in velocity range $[-3 \text{ km/s},3\text{ km/s}]$ with respect to the rest frame of source.
For all of our simulated emission, we assume local thermodynamic equilibrium (LTE) and set $T = 10\text{ K}$.
A total number of $10^5$ unpolarized background photons are initialized in each run, which are ray-traced to a $256^2$ detector.
For simulations of our \texttt{lmde} model, we place the detector at a distance $d = 150\text{ pc}$ from the box center, and for simulations of our \texttt{MSEnv} model we use $d = 1\text{ kpc}$.
These values are based on typical distances to nearby regions of low-mass star formation (e.g., the Taurus molecular cloud) and massive star formation, respectively.

For a locally uniform magnetic field that is weak enough such that any Zeeman splitting is unresolved \citep{crutcher1993}, the following relationship applies: 
\begin{align}
    V = \frac{dI}{d\nu} z_B B_{\rm LOS}\,, \label{eq:zeemaneq}
\end{align}
where $B_{\rm LOS}$ is the line-of-sight component of the magnetic field.
Using this relationship, we may compute the line-of-sight magnetic field strength required (under uniform magnetic field conditions) to achieve the nominal ALMA percentage polarization limit of 1.8\%.
For this, we ran a sample POLARIS simulation with a uniform magnetic field (oriented along the line-of-sight) in a box with uniform gas density.
We then calculated $dI/d\nu$ from the Stokes $I$ profile and found its maximum value, $(dI/d\nu)_{\rm max}$, as well as the maximum value of the Stokes $V$ profile, $V_{\rm max}$.
Under uniform conditions, the required line-of-sight magnetic field for 1.8\% is then
\begin{align}
    B_{\rm LOS} = 0.018 \times V_{\rm max} (\frac{dI}{d\nu})_{\rm max}^{-1} (z_B)^{-1}\,, \label{eq:zeemaneq_limit}
\end{align}
which yields a value of $B_{\rm LOS} \approx 3.1\text{ mG}$ in our case.
Throughout this work, we refer to this 3.1 mG estimate (in addition to the 1.8\% observational limit) as guidance for assessing potentially detectable magnetic field conditions in our simulations.
We also sometimes refer to a hypothetical factor-of-ten improved limit at 0.18\% and 0.31 mG, respectively.
This exercise is undertaken to evaluate the potentially improved utility of conducting Zeeman observations with a next generation circular polarization instrument of the future.

In Table \ref{table:3dperc}, we report the fraction of cells in each of our grids with magnetic field component values above 3.1 mG and 0.31 mG.

\begin{table}
\caption{Percentage of 3D cells on our re-projected fixed grids that have magnetic field component values above 3.1 mG and 0.31 mG. In addition to the full domain, we also perform computations that only include cells within a given radius; the chosen radii are the same we use for cuts in our polarization analysis in Section \ref{ssec:results_pp}. 
For model \texttt{MSEnv}, we also report values when only the cells within our CN enhanced shell are considered. 
\label{table:3dperc}}
\tabcolsep=0.12cm
\begin{tabular}{lc|cccc|cccc}
\hline
 & &  \multicolumn{8}{|c|}{Percentage of cells above... } \\ 
Model &  Cutoff & \multicolumn{4}{|c|}{3.1 mG} & \multicolumn{4}{|c|}{0.31 mG} \\ 
\cmidrule(lr){3-6}\cmidrule(lr){7-10}
& & $B_x$ & $B_y$ & $B_z$ & $B_{\rm tot}$ & $B_x$ & $B_y$ & $B_z$ & $B_{\rm tot}$ \\
\hline
\texttt{lmde} & - & 44.8 & 61.1 & 49.1 & 73.3 & 93.4 & 96.9 & 96.0 & 99.9 \\
 & $R < 75\text{ au}$ & 60.9 & 77.9 & 70.6 & 98.0 & 95.8 & 98.4 & 97.3 & 100 \\
\hline
\texttt{MSEnv} & - & 24.2 & 31.4 & 43.9 & 42.2 & 82.1 & 82.6 & 92.1 & 92.3 \\
 & $ R < 2500\text{ au}$ & 57.2 & 62.9 & 72.1 & 81.2 & 93.3 & 91.4 & 94.0 & 96.6 \\
 & In shell & 70.0 & 68.6 & 72.2 & 86.5 & 97.2 & 97.3 & 97.5 & 99.9 \\
\hline
\end{tabular}
\end{table}

\section{Results} \label{sec:results}
Presented in this section are our circular polarization results for both the \texttt{lmde} and \texttt{MSEnv} models.
We include integrated emission maps (Section \ref{ssec:results_maps}) and computations of the local polarization percentage in locations across different cuts of the simulated observer-space (Section \ref{ssec:results_pp}).
We also use 2-dimensional histograms to relate our polarization data to local magnetic field information (Section \ref{ssec:results_2dhist}).

\subsection{Maps} \label{ssec:results_maps}
\begin{figure*}
\centering
\includegraphics[width=\textwidth]{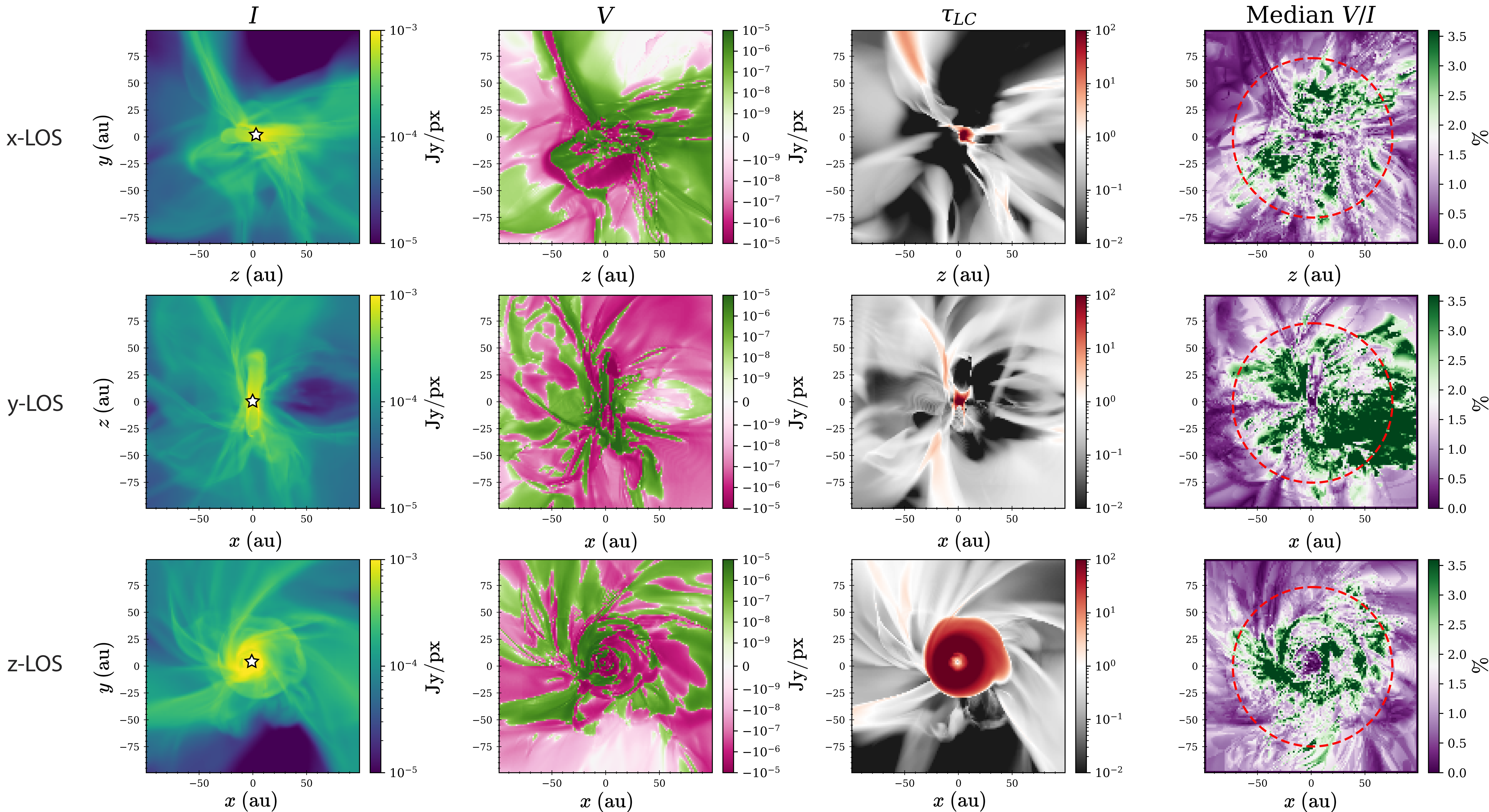}
\caption{Velocity-integrated Stokes $I$ and Stokes $V$, line center optical depth $\tau_{LC}$, and derived polarization quantity ``Median $V/I$'' for each Cartesian view of the \texttt{lmde} model. The circle annotated on the last column corresponds to $R = 75\text{ au}$.} 
\label{fig:yiobs}
\end{figure*}

\begin{figure*}
\centering
\includegraphics[width=\textwidth]{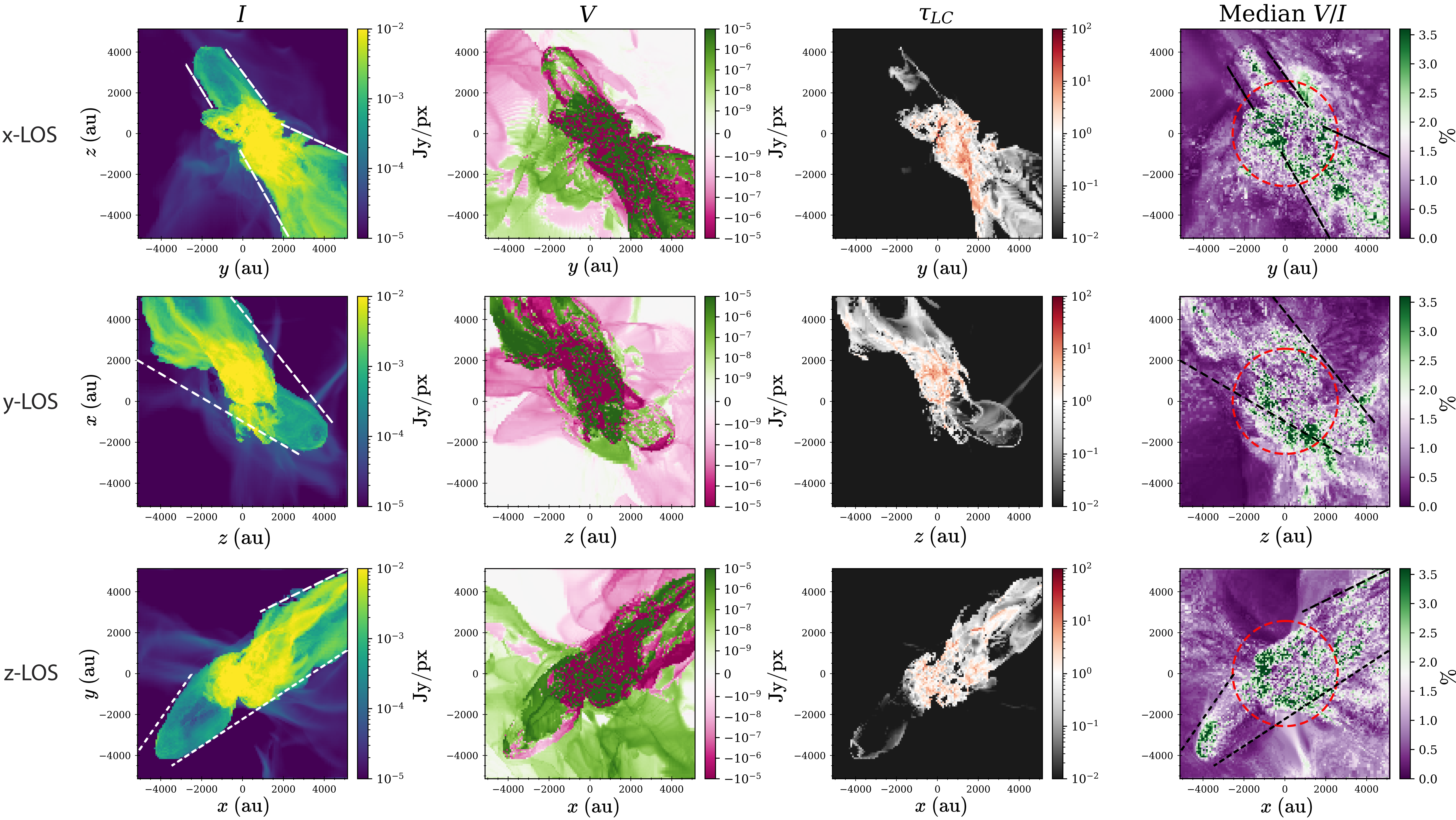}
\caption{Same as Figure \ref{fig:yiobs}, but now for the \texttt{MSEnv} model. The circle annotated on the last column corresponds to $R = 2500\text{ au}$.
We also draw additional reference lines on the Stokes $I$ panels (in white) and on the Median $V/I$ panels (in black). These sets of lines are co-located with each other and are drawn in by hand to guide visual inspection of the polarization in the CN enhanced shell and its nearby surroundings.} 
\label{fig:cyobs}
\end{figure*}

In Figure \ref{fig:yiobs} and Figure \ref{fig:cyobs} we show maps (for each of the $x-$, $y-$, and $z-$sightlines) of the observable data obtained from our POLARIS simulations of model \texttt{lmde} and model \texttt{MSEnv}, respectively.
Included are the velocity-integrated Stokes $I$ and Stokes $V$, the CN optical depth at line-center ($\tau_{LC}$), and a derived percentage polarization quantity that we refer to as the ``Median $V/I$''. 
To calculate this value, we bin down our data from 169 channels to 13 channels, resulting in a velocity resolution of 0.46 km/s that is comparable to that which is typical of Zeeman observations with ALMA. 
For each pixel in the observer space, we then calculate the ratio of the Stokes $V$ to the Stokes $I$ in each of the 13 re-binned channels.
The ``Median $V/I$'' for a given pixel is then the 7th highest of these 13 values.
Functionally, this parameter produces similar percentage polarization values to comparing the peaks of the Stokes $V$ versus Stokes $I$, while also imposing the requirement that the majority of channels must have detectable emission (which is a soft criterion for being able to reasonably estimate the local line-of-sight magnetic field strength, e.g., by applying equation \ref{eq:zeemaneq}).

Generally, the maps for both models show that the percentage polarization is maximized in at an intermediate distance from the central protostar.
In model \texttt{MSEnv}, for all three viewing angles there is a low polarization ``hole'' near the star (within $R \lesssim 500\text{ au}$) where the line-center optical depth is high ($\tau_{LC} > 1$).
Moreover, the locations of high polarization are spatially correlated with the location of the enhanced CN outflow shell defined by our tracer particle constraint. 

Our results from model \texttt{lmde} also reveal that there is polarization sub-structure when we focus in on the $R \lesssim 100\text{ au}$ scale near the disk. 
In this case, however, rather than a roughly spherically symmetrical low polarization ``hole,'' the low polarization region is defined by the shape of the (optically thick) disk.
For the edge-on views (i.e., the $x-$ and $y-$sightlines), the percentage polarization is low along the disk midplane, with pixels that exceed $1.8\%$ polarization fraction mainly occurring between $5-75\text{ au}$ above or below the midplane. 
In the $y$ line-of-sight, there is also clearly a region of high polarization that spatially overlaps with the location of a disk wind ($z \gtrsim 10\text{ au}$ along the disk axis).
Looking face-on at the disk ($z$ line-of-sight), there is a circular low polarization region near disk center, again corresponding to where $\tau_{LC} > 1$.
From this view, pixels in excess of $1.8\%$ tend to occur in a radial band $R \sim [30\text{ au},50\text{ au}]$ surrounding the edge of the optically thick disk. 

The polarization percentage is also, unsurprisingly, low in the outskirts of the box domains, in the areas far beyond the CN enhanced shell or disk edge, for models \texttt{MSEnv} and \texttt{lmde} respectively, where the magnetic field tends to be weaker in general.

\subsection{Percentage Polarization Statistics} \label{ssec:results_pp}
As a corollary to Table \ref{table:3dperc}, where we listed the fraction of 3D cells with magnetic field strengths above 3.1 mG and 0.31 mG, in Table \ref{table:percentages} we calculate the fraction of pixels from our simulated emission maps that have percentage polarization (``Median $V/I$'') values above 1.8\% and 0.18\%.
For both simulation types we perform these computations for the full observer-space, as well as for cuts that restrict the domain to more favorable sub-regions (from a percentage polarization perspective). 
The locations of these cuts are also noted in Table \ref{table:percentages}.

For the low mass disk model \texttt{lmde}, we find that across the whole observer space the fraction of cells with fractional polarization above 1.8\% is between $\sim$$20-40\%$ depending on viewing geometry.
Interestingly, the $y$ line-of-sight in this case has a higher fractional polarization value largely due to a swath of high polarization pixels along a disk wind.
When Cut 1 is applied (i.e., the highest column density cells corresponding to the high $\tau$ disk are removed from consideration), there is effectively no change in the fraction of cells with percentage polarization above 1.8\%.
This is sensible, since the disk only accounts for a small area of the observer space.
Visually, as we see in Figure \ref{fig:yiobs}, it is clear however that the inner disk has quite low polarization.
When we add in the restriction to limit the probed space to $ R < 75\text{ au}$ (Cut 2), the percentage polarization increases by factors of $\sim$2, 1.5, and 2 for the $x-$, $y-$, and $z-$lines-of-sight, respectively.
This calculation aligns well with what is suggested in the maps; that the regions of high polarization tend to lie spatially in an intermediate range just beyond the disk edge, where the magnetic field is relatively strong and the optical depth is not too high.

The results for the massive star case are qualitatively similar, in that the fraction of pixels above 1.8\% polarization is highest when only an annulus is considered.
For the whole domain, the percentage of high polarization ($>$$1.8\%$) pixels is about $\sim$15\% for all three Cartesian sightlines.
This value increases by about a factor of 2 up to $\sim$30-35\% when only pixels within 2500\text{ au} of the central star are included (Cut 1).
We also define a cut (Cut 2) that additionally removes the polarization ``hole'' in the $x$ line-of-sight view.
The location of this mask is set manually to be centered on the lowest polarization pixel in that region, with a radius of 900 au.
Removing the hole from consideration increases the fraction of pixels with percentage polarization above 1.8\% from 35\% to 41\% for the $x$ line-of-sight.

Finally, it is also worth noting that for both models and all cuts, the fraction of cells with percentage polarization above 0.18\% always exceeds 90\%.
This suggests that a hypothetical factor of ten improvement in sensitivity to the fractional circular polarization would result in the vast majority of the envelopes becoming accessible with Zeeman observations.

\begin{table*}
\caption{Computations of the fraction of pixels in each of our simulated emission maps with percentage polarization values above 1.8\% and 0.18\%. We report values for the entire box domain for each of the Cartesian lines-of-sight, as well as for some selected cuts within the observer-space. \label{table:percentages}
}
\begin{tabular}{lcccccc}
\hline
 & & & & \multicolumn{2}{|c|}{Percentage of pixels above... } & \\ 
Model & LOS & Inner cutoff & Outer cutoff & 1.8\% & 0.18\% & Notes\\  

\hline

\texttt{lmde} & $x$ & - & - & 21.8 & 94.5 & Full observer-space\\
& $y$ & - & - & 40.9 & 95.6 \\
& $z$ & - & - & 24.2 & 95.9 \\
& $x$ & $N > 60\text{ g cm}^{-2}$ & - & 21.9 & 94.6 & Cut 1\\
& $y$ & $N > 60\text{ g cm}^{-2}$ & - & 41.3 & 95.7 \\
& $z$ & $N > 60\text{ g cm}^{-2}$ & - & 23.1 & 96.1 \\
& $x$ & $N > 60\text{ g cm}^{-2}$ & $R < 75\text{ au}$  & 39.9 & 99.8 & Cut 2 \\
& $y$ & $N > 60\text{ g cm}^{-2}$ & $R < 75\text{ au}$  & 62.2 & 99.8 \\
& $z$ & $N > 60\text{ g cm}^{-2}$ & $R < 75\text{ au}$  & 45.3 & 99.6 \\
\hline
\texttt{MSEnv} & $x$ & - & - &  17.7 & 96.4 & Full observer-space \\
& $y$ & - & - & 17.6 & 89.2 \\
& $z$ & - & - & 14.1 & 96.6 \\
& $x$ & - & $R < 2500\text{ au}$ & 34.7 & 97.9 & Cut 1 \\
& $y$ & - & $R < 2500\text{ au}$ & 30.6 & 97.9 \\
& $z$ & - & $R < 2500\text{ au}$ & 38.8 & 97.9 \\
& $x$ & $R > 900\text{ au}$ & $R < 2500\text{ au}$ & 41.3 & 99.5 & Cut 2\\

\hline
\end{tabular}
\end{table*}

\subsection{2-Dimensional Histograms} \label{ssec:results_2dhist}

\begin{figure*}
\centering
\includegraphics[width=\textwidth]{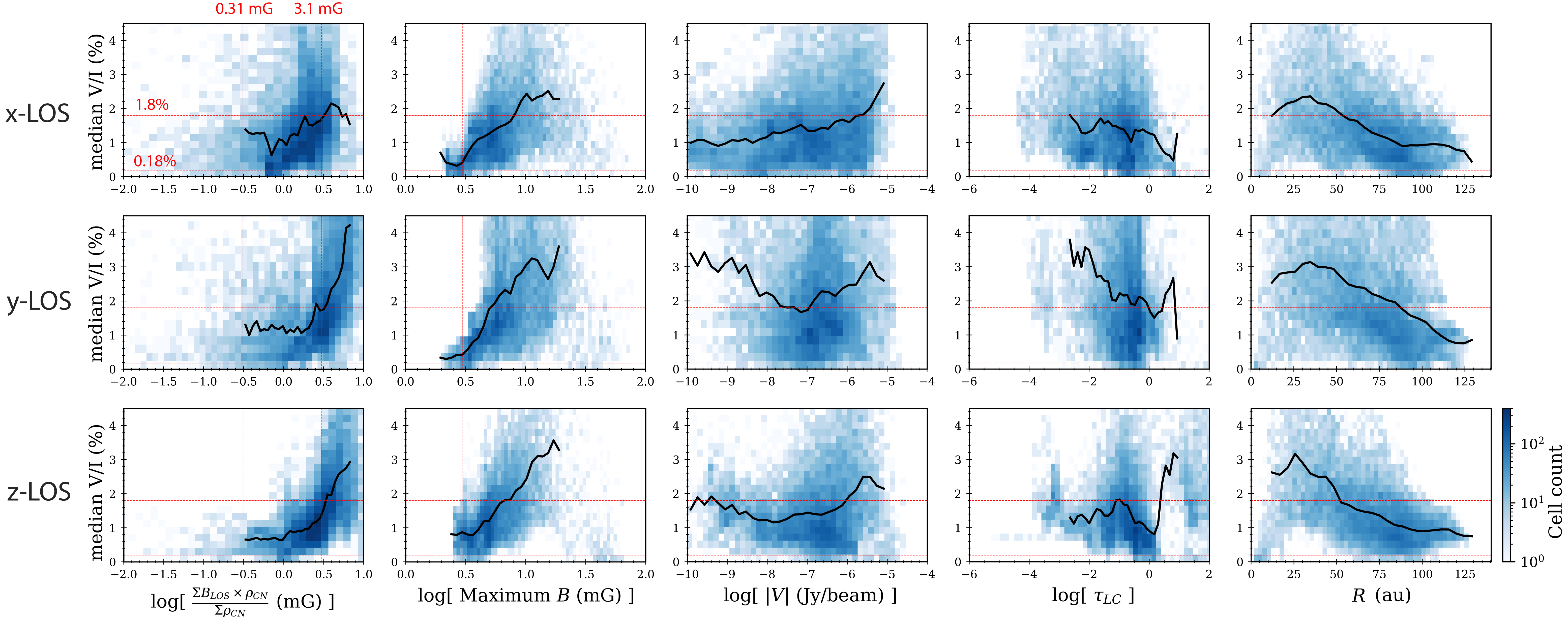}
\caption{Model \texttt{lmde} 2-dimensional histograms (for each Cartesian line-of-sight) comparing the median $V/I$ quantity (vertical axis) in each pixel with several other quantities derived from the same pixel. 
For each pixel these quantities include, from left-to-right, the density weighted average line-of-sight magnetic field strength, the maximum magnetic field strength in any 3D cell along that sightline, the absolute value of the velocity-integrated Stokes $V$ signal, the line-center optical depth, and the distance (in observer space) from the central sink particle.}
\label{fig:2dhist_sm}
\end{figure*}

\begin{figure*}
\centering
\includegraphics[width=\textwidth]{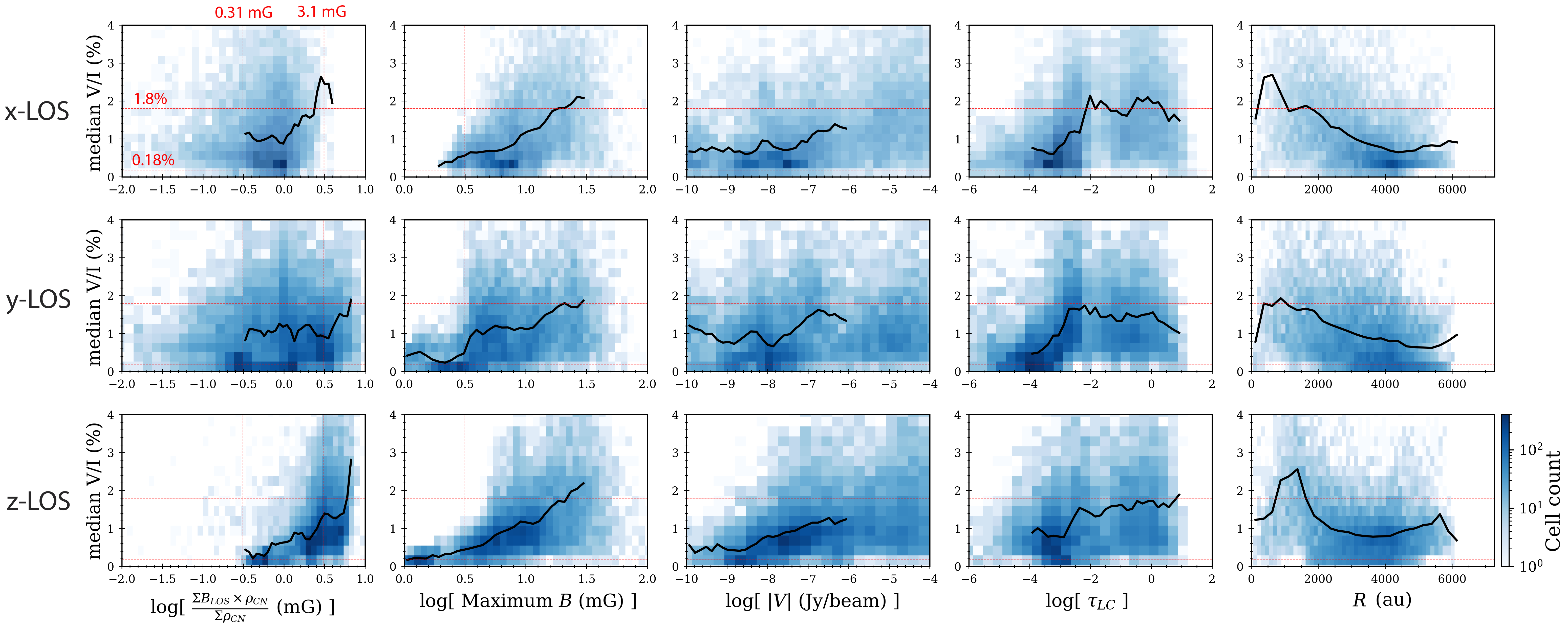}
\caption{Same as Figure \ref{fig:2dhist_sm}, but now for model \texttt{MSEnv}.}
\label{fig:2dhist_hmw}
\end{figure*}

Here we relate our percentage polarization $V/I$ results to the underlying magnetic field information as well as other observable quantities of interest (i.e., Stokes $V$ and CN $\tau_{LC}$).
Two-dimensional histogram plots for models \texttt{lmde} and \texttt{MSEnv} are presented in Figure \ref{fig:2dhist_sm} and Figure \ref{fig:2dhist_hmw}, respectively.

Each plot has significant scatter, suggesting that high polarization can in principle be obtained under a wide variety of envelope conditions.
However, there are also some trends apparent in our data that show some characteristics are more favorable than others. 
As expected, Median $V/I$ scales positively with the density weighted average line-of-sight magnetic field strength (first column of Figures \ref{fig:2dhist_sm} and \ref{fig:2dhist_hmw}).
Moreover, for both simulations the associated mean trend line tends to pass through a polarization percentage of 1.8\% at roughly 3.1 mG.
This suggests that this equivalent value for a uniform magnetic field we calculated in Section \ref{sec:zeeman} is a useful guide for interpreting envelope Zeeman signatures.
That is to say, it is reasonable to expect that robust detections with ALMA should require average magnetic fields strengths $\gtrsim$ 3 mG.
Note, however, that this value applies only in envelope regions like those we simulate here, where the magnetic field is mostly uniform or only moderately tangled.
In regions with more complex (i.e., tangled or wound-up) magnetic field geometry, larger average field strengths will be required to achieve the same polarization percentage. 
It is also the case that the maximum magnetic field strength (along the line-of-sight) scales positively with median $V/I$ (second column of Figures \ref{fig:2dhist_sm} and \ref{fig:2dhist_hmw}).
For pixels with percentage polarization greater than 1.8\%, this maximum value always exceeds 3.1 mG and can reach as high as $\sim$10-30 mG, which is consistent with the expectation that the Zeeman-inferred field strength is a lower limit to the maximum field strength found along the line-of-sight. 

Another notable trend is that pixels with strong velocity-integrated $|V|$ signals tend to also have large $V/I$.
This is especially true at large $|V|$.
This result is convenient for observers, because it means that locations with high percentage polarization will also be the most likely to have a detectable Stokes $V$ signal.

For both model types, there is also a turnover in percentage polarization at an intermediate radius.
The peaks occur at approximately $\sim$30 au for model \texttt{lmde} and $\sim$1500 au for model \texttt{MSEnv}.
As we touched on in Section \ref{ssec:results_maps}, visual inspection of the maps of the observables reveal that these radii correspond approximately to the edge of the optically thick disk and the polarization ``hole'' for the two model types, respectively.

Assessing percentage polarization trends with line-center optical depth is less clear.
Overall, there is a wide range of $\tau_{LC}$ values that can produce potentially detectable percentage polarization levels in both envelopes.
In model \texttt{lmde} there tends to be a moderate anti-correlation between optical depth and $V/I$, especially in the opacity regime where the majority of the pixels tend to be located ($\log{[\tau_{LC}]} \sim [-2,0]$).
For the face-on case, there is also a local maximum (in the mean curve) around $\tau_{LC} \sim 0.1$, which lends some credence to the idea the intermediate optical depth may sometimes be favorable.
For model \texttt{MSEnv} there tends to be low polarization at very small $\tau_{LC} \sim -4$.
This is due to low-density ambient region far outside the CN enhanced shell.
Within and near the shell (where $\log{[\tau_{LC}]} \sim [-2,1]$) the percentage polarization tends to be close to 1.8\% with very little discernible trend, except for a downturn at the highest opacities ($\log{[\tau_{LC}]} \gtrsim 0$) for the $x$ and $y$ lines-of-sight.
This modest anti-correlation corresponds to the polarization ``hole,'' which is evidently not as visible in the $z-$view.

One further caveat that we should note about optical depth is that our simulations are tailored toward the optically thin scenario; we chose our CN abundances such that we would generally have $\tau_{LC} < 1$ throughout the envelopes of our modeled sources.
This is in part because we expect modeled optically thin emission to be more directly comparable to observations than optically thick emission.
In practice, optically thick regions are subject to additional line effects that make drawing comparisons to modeled emission more tenuous, such as continuum over subtraction and absorption by resolved-out, cold outer envelope material.

\section{Discussion} \label{sec:disc}

Generally, the results of our simulations suggest that observable $V/I$ signals (with e.g., ALMA) are available in both massive and low-mass protostellar envelopes.
Therefore, probing magnetic fields in these environment with the Zeeman effect is a tractable goal.
Our results also show, however, that there is significant variation in percentage polarization across the envelopes; that is to say, some sight-lines are more favorable than others.

In this discussion, we investigate more deeply some factors that can influence whether a given location is a good candidate to have high percentage polarization: 
%line-of-sight magnetic field structure (Section \ref{ssec:disc_los}), 
optical depth (Section \ref{ssec:disc_depth}), CN enhancement on a wind-blown shell (Section \ref{ssec:disc_cnenh}), and disk inclination (Section \ref{ssec:disc_inc}).
We then highlight in Section \ref{ssec:disc_obscav} some additional technical factors that observers should be mindful could also affect the percentage polarization results.
Finally, for our \texttt{lmde} simulation we provide example lines (Stokes $I$ and Stokes $V$ profiles) from a selected envelope location to give a sense of typical line morphology and linewidth in this model (Section \ref{ssec:disc_lines}). We also compare the magnitude of the integrated flux found in this simulation to TMC-1, an example Class I source that is a good candidate for inner envelope Zeeman observations.

\subsection{Optical Depth} \label{ssec:disc_depth}
\begin{figure*}
\centering
\includegraphics[width=\textwidth]{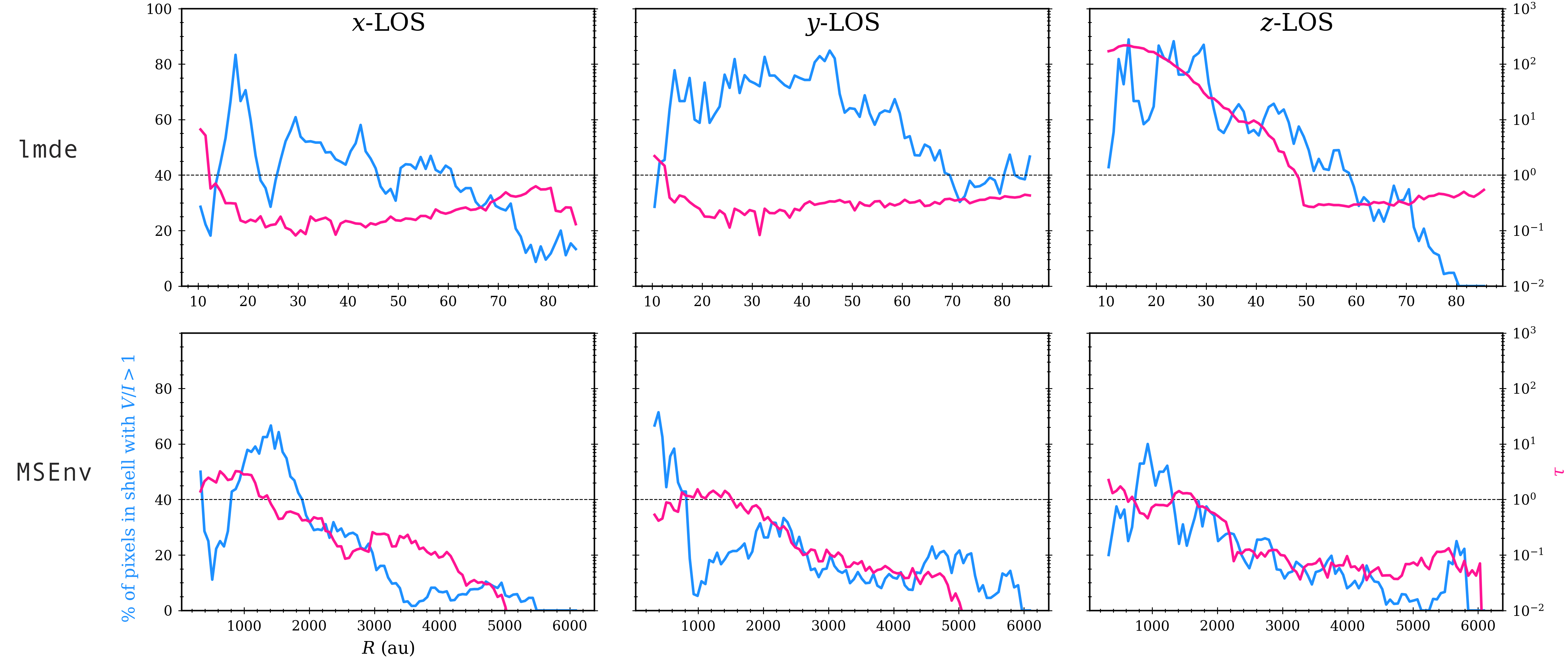}
\caption{Percentage of pixels with median polarization percentage above 1.8\% (blue curve) and average (mean) optical depth (pink curve) calculated in thin radial shells, computed for both of our models as observed along each Cartesian line-of-sight. A horizontal line corresponding to a CN line-center optical depth of unity is also annotated on each plot.} 
\label{fig:shells}
\end{figure*}

From our 2-dimensional histogram analysis in Section \ref{ssec:results_2dhist} we concluded that global trends in $V/I$ versus line-center optical depth are not especially clear for either model type.
Here, we re-visit the topic with an alternative approach.
In Figure \ref{fig:shells} we plot the radial profiles of two quantities, the fraction of cells with polarization percentage above 1.8\% and the average $\tau_{LC}$, within thin annuli of radial size $[R-\Delta r/2, R+\Delta r/2]$. For model \texttt{lmde} we set $\Delta r = 6$ au and for model \texttt{MSEnv} we set $\Delta r = 300$ au.

Our plots for the model \texttt{lmde} show different trends depending on line-of-sight.
For the edge-on disk views ($x-$ and $y-$ lines of sight), the percentage of pixels above 1.8\% polarization percentage is relatively low at small radius $(R \lesssim 15\text{ au}$) then peaks around $R = 20-50\text{ au}$.
In this radial range the average optical depth hovers around $\tau_{LC} \sim 0.2 - 0.4$.
Visual inspection of the maps (top two rows of Fig. \ref{fig:yiobs}) shows us that these results are due to high polarization swaths above and below the disk midplane.
The face-on view of model \texttt{lmde}, by contrast, shows that the annuli with the highest fraction of high polarization cells also have the highest optical depth.
While it is certainly the case that some pixels in the optically thick disk have high polarization, taken at face value this result is perhaps somewhat misleading.
Looking at Figure \ref{fig:yiobs} again (bottom row), we can see that most of the pixels with median $V/I > 1.8\%$ are either in the transition region near the disk edge (where $\tau$ is rapidly decreasing) or further away.
Furthermore, in the face-on view of the disk there are opposing radial asymmetries in both the optical depth map and the $V/I$ map in the range $R \sim 30-50\text{ au}$.
Particularly, there is a lobe of high $\tau_{LC}$ at positive $x$, whereas the high $V/I$ extension is at negative $x$ where the CN optical depth is actually lower, at around $\tau_{LC} \sim 0.1-1$.

For model \texttt{MSEnv}, all three lines-of-sight show similar trends.
Though at small $R$ there is some difference in where the high polarization regions lie as a function of radius (this is due to slight differences in the location of the central polarization ``hole'' for each projection), within $R \lesssim 2500\text{ au}$ it is generally the case that peaks of the fraction of pixels with $V/I > 1.8\%$ are accompanied by dips in the average line center optical depth (to $\tau_{LC} \lesssim 1$).
Outside $R = 2500\text {au}$, both quantities gradually taper off toward the outskirts of the envelope. 

\subsection{CN Enhancement} \label{ssec:disc_cnenh}
In our simulations of model \texttt{MSEnv} we included a non-uniform abundance prescription, with 3D cells lying within a prescribed shell region (see Fig. \ref{fig:shell}) being supplied with a CN abudance a factor of 1000 larger ($X_{\rm shell} = 10^{-9}$) than the surrounding ambient material ($X_{\rm shell} = 10^{-12}$).
The results of our radiative transfer simulations for this model therefore give us some opportunity to comment on how the locations of CN enhanced regions, particularly those which may be driven by UV irradiation on the cavity wall from a protostellar wind, can inform observational choices when designing a Zeeman experiment.

In our plot of the observable data from the massive star simulation (Figure \ref{fig:cyobs}), we annotate reference lines on the Stokes $I$ maps (white dashes) and median $V/I$ maps (black dashes). 
These lines are co-located and drawn in by hand strictly for visual reference.
From inspection of the maps it is clear that the regions of high polarization are well-correlated with the CN enhanced shell.
Particularly high polarization occurs near the shell edges surrounding the central polarization ```hole'' or in its lower-intensity (as measured by Stokes $I$ value) fringes.
For example, the blown out ``tip'' of the shell located at $(x,y,z) \sim (-3000\text{ au}, -1000\text{ au}, 3000\text{ au})$ contains many high polarization pixels.
These results suggest that interfaces between outflow structures and ambient gas (e.g., due to outflow-swept shells) are prime areas to target, due both to favorable CN abundances and relatively low to intermediate optical depth compared to (low polarization) central regions. 

\subsection{Disk Inclination} \label{ssec:disc_inc}
So far in this work we have only considered simulations viewed from the Cartesian lines-of-sight (with respect to the frame of our MHD simulations). 
While this may be sufficient to obtain a good understanding of the \texttt{MSEnv} model, which is more-or-less spherically symmetric in its interior region, our results for the 
\texttt{lmde} model show significant contrast in polarization map morphology between the face-on and edge-on disk views.

\begin{figure*}
\centering
\includegraphics[width=\textwidth]{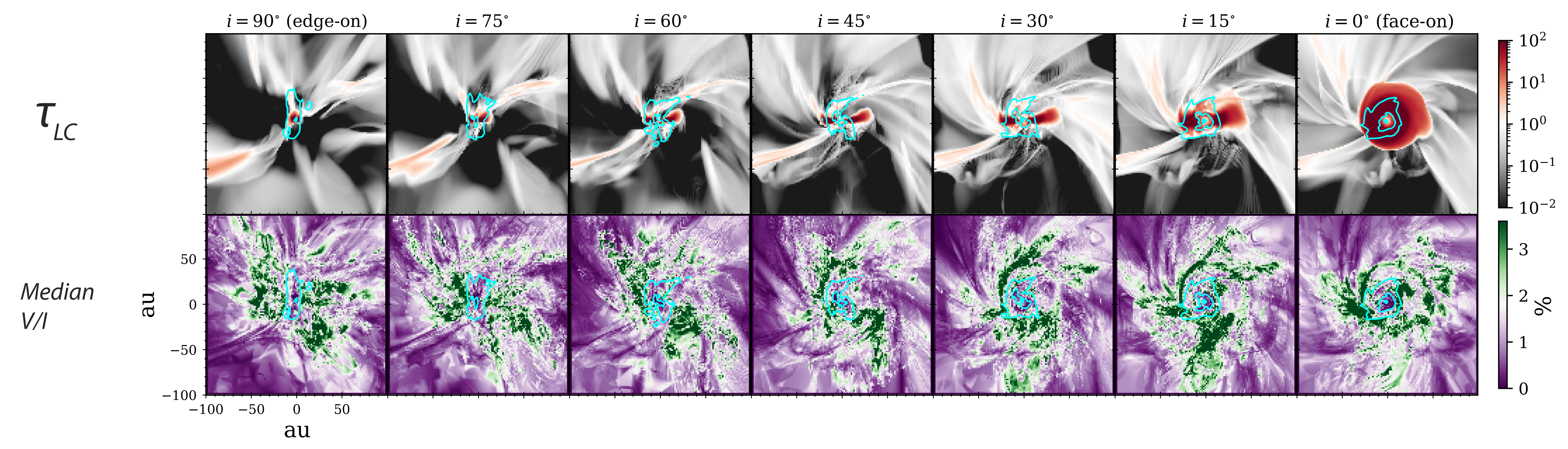}
\caption{Line-center optical depth and median $V/I$ calculated for several intermediate inclination views of model \texttt{lmde}. Over-plotted on each panel in cyan are contours corresponding to $5 \times 10^{-4}$ and $10^{-3}$ Jy/px, which trace out the brightest parts of the central disk. For all viewing angles, these brightest parts of the disk tend to have low polarization and are immediately surrounded by high polarization regions. For views that are close to edge-on ($\sim$60$^{\circ}$-90$^{\circ}$) the high polarization regions tend to be above and below the disk midplane, and for views close to face-on ($\sim$0-30$^{\circ}$) the high polarization regions roughly form a ring around the optically thick disk.)} 
\label{fig:disk_inc}
\end{figure*}

Figure \ref{fig:disk_inc} presents percentage polarization results for a series of intermediate disk inclination views of our low mass protostar simulation.
The general picture is similar for all viewing orientations; there is a central (highest intensity in Stokes $I$ and highest optical depth) region that is low in polarization, and it is surrounded by a relatively high polarization envelope.
The shapes of the regions with large $V/I$ vary as a function of inclination.
As the angle is adjusted from $i = 90^{\circ}$ (edge-on) to $i = 0^{\circ}$ (face-on), the high polarization swaths gradually progress from appearing shaped like blocks above and below the disk midplane into a ring near the disk edge.
From this experiment we can see that intermediate inclination low mass protostellar envelopes are potential targets for Zeeman experiments, as they also contain many high polarization pixels.
They just perhaps have more complex morphology, with the high polarization regions appearing as some combination of ``block-like'' and ``ring-like.'' 

\subsection{Additional Observational Considerations} \label{ssec:disc_obscav}

In Sections \ref{ssec:disc_obscav_bc} and \ref{ssec:disc_obscav_st} below, we briefly assess the impact beam convolution and stacking of sub-transitions can have on our observational results by considering a few case studies.

\subsubsection{Beam Convolution} \label{ssec:disc_obscav_bc}
We test the effect of applying 0.5$^{\prime \prime}$ FWHM beam convolution to the $x$ line-of-sight view of our \texttt{MSEnv} model.
For this experiment we bin our 169 channels of Stokes $I$ and $V$ data to 13 channels (corresponding to an observed velocity resolution of 0.46 km/s) as usual, then at that stage apply a Gaussian filter to each of those 13 channels. 
We then use those data to compute velocity-integrated $I$ and $V$, line-center CN optical depth, and the median $V/I$.

\begin{figure*}
\centering
\includegraphics[width=\textwidth]{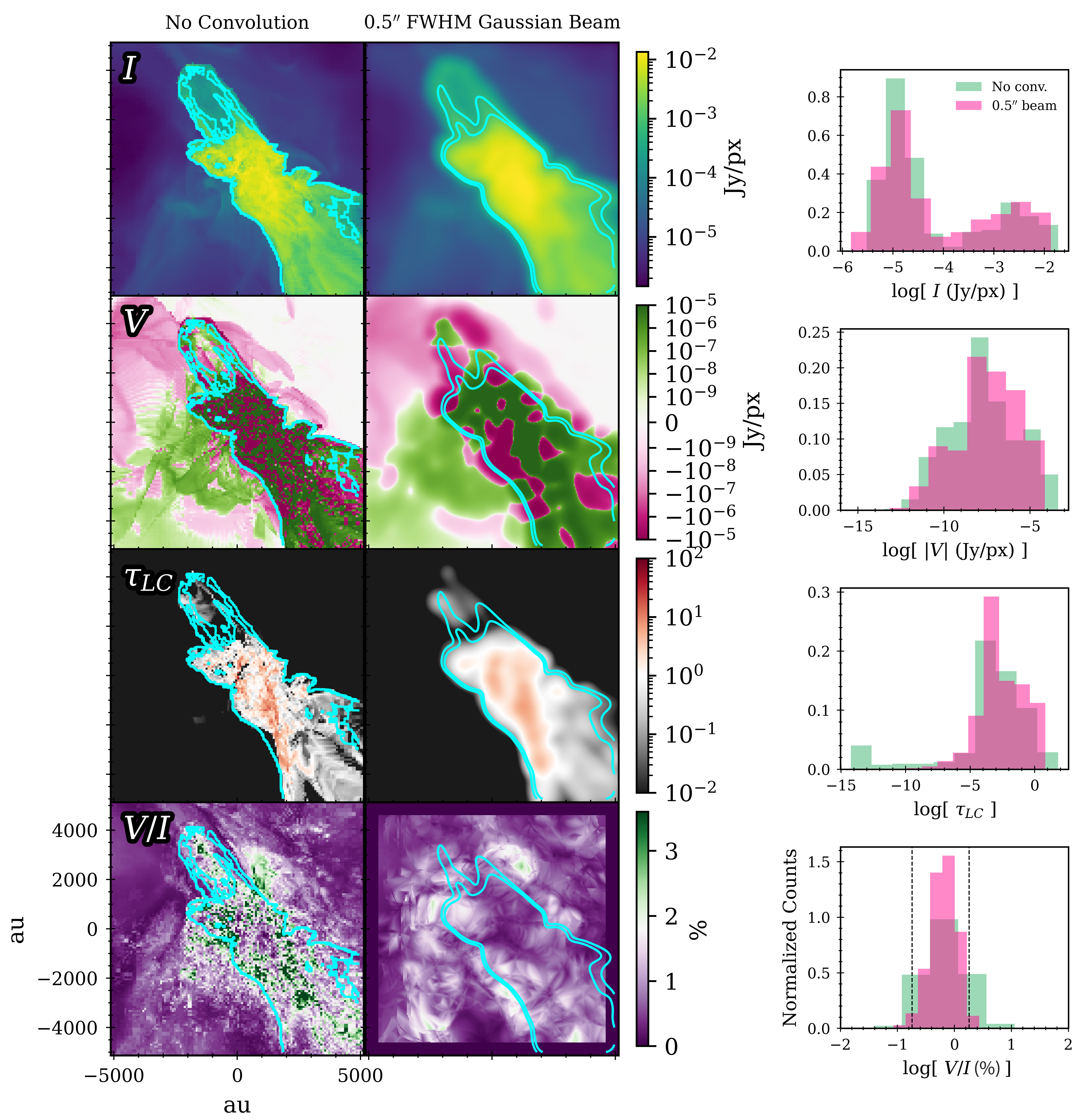}
\caption{\textbf{\textit{Left:}} Comparison of velocity-integrated Stokes $I$ and $V$, line-center optical depth, and median polarization percentage between pixel resolution and $\theta_{\rm FWHM} = 0.5^{\prime \prime}$ beam convolved cases, for model \texttt{MSEnv} as viewed from the $x$ line-of-sight. The cyan contours correspond to Stokes $I$ values of $5 \times 10^{-4}$ and $10^{-3}$ Jy/px. \textbf{\textit{Right:}} Histogram plots comparing the distribution of each observable for the unconvolved versus convolved cases. The main effect of applying beam convolution is that the $V/I$ distribution narrows, resulting in fewer pixels having percentage polarization values above $1.8\%$. The distribution remained peaked at roughly the same value, however ($V/I\sim0.8\%$).} 
\label{fig:conv}
\end{figure*}

A side-by-side comparison of our high resolution maps (i.e., with no beam convolution) and our $\theta_{\rm FWHM} = 0.5^{\prime \prime}$ maps are presented in Figure \ref{fig:conv}. 
We also include histograms of each observable to compare their values in aggregate.
In addition to the obvious morphological changes from Gaussian smoothing, the main effect is in the $V/I$ map.
Clearly, the values of the pixels with the highest percentage polarization are significantly reduced.
This is an expected result, as beam convolution effectively averages adjacent pixels.
Therefore, any high polarization regions that are smaller than the beam size will have some contribution from low polarization pixels post-convolution.
This effect is captured in the histograms as well.
The overall Stokes $I$ and $V$ distributions are similar in shape, but the convoluted $V/I$ distribution is notably narrower than the unconvolved version.
The percentage of pixels above 1.8\% fractional polarization is $\sim$15\% for the unconvolved map and $\sim$3\% for the convolved map.
Meanwhile, the percentage of pixels above 0.18\% fractional polarization is $\sim$94\% for the unconvolved map and $\sim$99\% for the convolved map.
It should also be noted that the two $V/I$ distributions peak at roughly the same value, $V/I \sim 0.8\%$.

These results suggest that either high sensitivity ($V/I\text{ limit} < 1.8\%$) or high resolution ($\theta_{\rm FWHM} < 0.5^{\prime \prime}$), or a combination of both, would be of great advantage for Zeeman experiments in this type of environment.

\subsubsection{Stacking Sub-Transitions} \label{ssec:disc_obscav_st}
The CN $J = 1 - 0$ molecular line comprises nine velocity-resolved hyperfine sub-transitions.
For the preceding sections of this work we only considered one representative transition, the 113.144 GHz component.
Here, we perform simulations of all seven\footnote{These are the seven ``main'' transitions in the CN $J = 1 - 0$, the other two have much lower intensity.} of the sub-components that are available by default in POLARIS.
We run these computations on the face-on ($z$ line-of-sight) view of our low mass disk envelope simulation, which serves as a good test case because it contains a wide variety of optical depth conditions.

\begin{figure*}
\centering
\includegraphics[width=\textwidth]{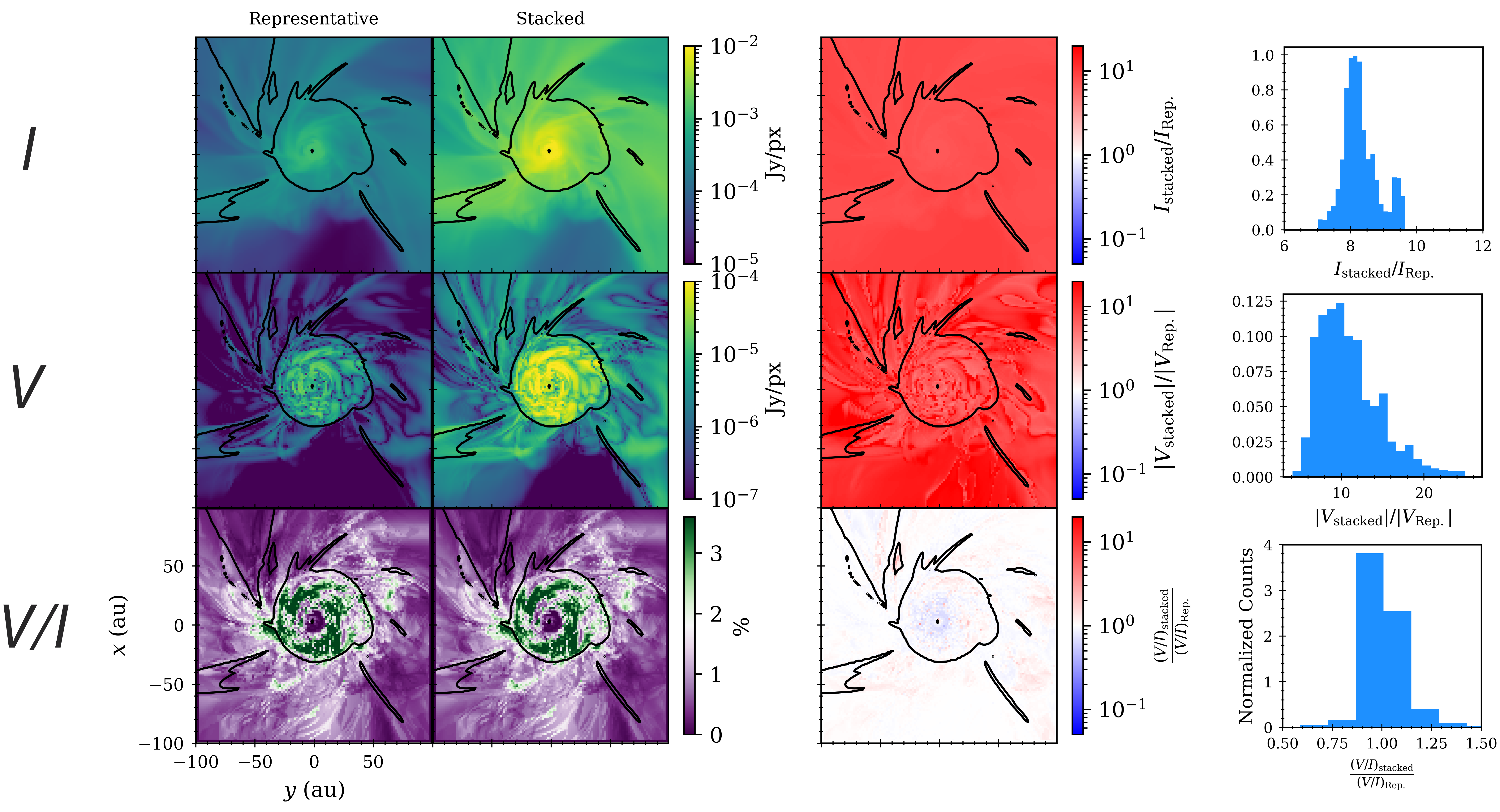}
\caption{\textbf{\textit{Left:}} Comparison of the velocity-integrated $I$ and $V$ signals and median polarization percentage calculated from the single 113.144 GHz representative transition versus those calculated after sub-transition stacking. Over-plotted contours correspond to $\tau_{LC} = 1$.
\textbf{\textit{Center}}: Pixel-by-pixel ratios of the observables for the stack versus the representative transition. \textbf{\textit{Right}}: Histogram plots of the same ratios. We generally find that the amount of signal boost gained from stacking has some modest pixel-by-pixel variation, with both distributions centered on a factor of $\sim$8.} 
\label{fig:stack}
\end{figure*}

We find that each of the seven components have nearly identical Stokes $I$, $V$, and $\tau$ morphology, so the main effect of stacking is a boosted gain in signal.
In Figure \ref{fig:stack} we compare velocity-integrated $I$ and $V$, and median $V/I$ maps obtained from the stacked transitions versus those from the representative transition.
We also compute the ratio between the stacked value and representative transition value for each of these quantities.
Stacking produces a (per pixel) boost in the $I$ signal by a factor of $\sim$7-10, with the distribution strongly peaked at roughly factor of 8.
The $V$ signal is boosted by roughly the same factor on average, but the distribution is a bit broader; some pixels reach as high as a factor of $\sim$25 brighter.
Finally, the median $V/I$ is about the same for the stacked data as it is for the single sub-transition.
Interestingly, the regions with the lowest values for this quantity [$(V/I)_{\rm stacked} / (V/I)_{\rm Rep.} \sim 0.8$] tend to be located within the optically thick disk, and the regions where stacking increases $V/I$ tend to be in the envelope.

\subsection{Example lines in model \texttt{lmde} and comparison of integrated Stokes $I$ flux to TMC-1} \label{ssec:disc_lines}

\begin{figure*}
\centering
\includegraphics[width=\textwidth]{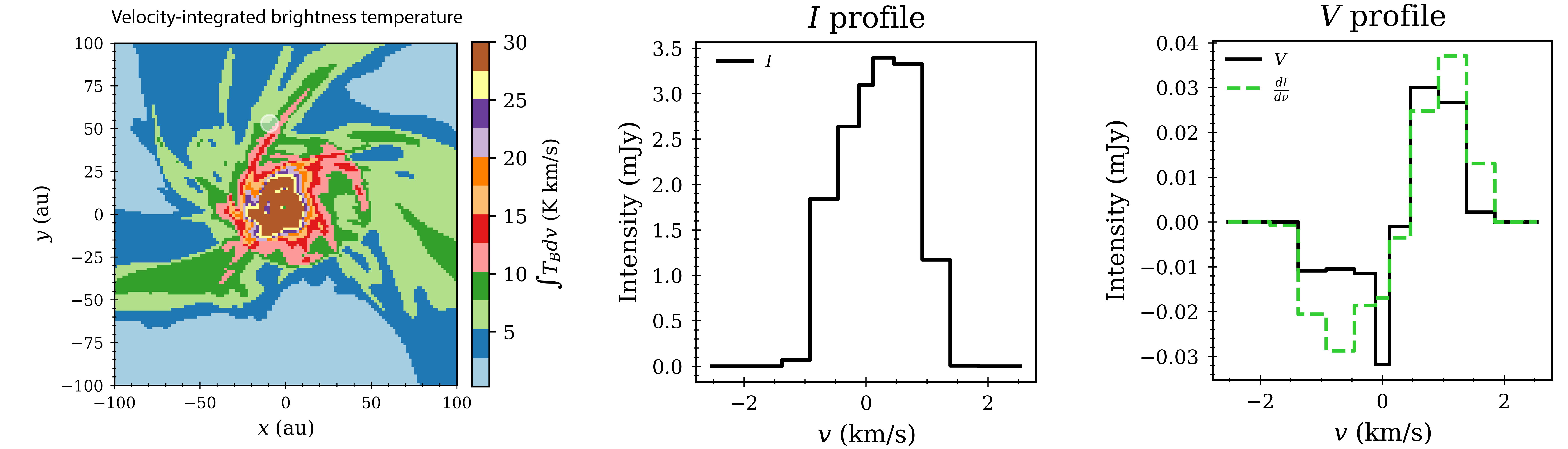}
\caption{\textbf{\textit{Left}}: Velocity-integrated brightness temperature for each pixel in the face-on view of our \texttt{lmde} model, expressed in K km s$^{-1}$ units. Also annotated on the plot is a translucent white circle, which represents an example beam location. \textbf{\textit{Middle and Right}}: Stokes $I$ and Stokes $V$ profiles calculated from the annotated example beam, to demonstrate the line morphology and linewidth for a typical envelope location. Also included on the $V$ profile plot is a fit curve using $dI/d\nu$. For this particular location, we obtain a polarization percentage of $\sim$$1\%$.}
\label{fig:lines_lmde}
\end{figure*}

From the perspective of an observer, it is also useful to get a sense of what the Stokes $I$ and $V$ line profiles look like in a typical envelope beam.
In Figure \ref{fig:lines_lmde}, we provide the velocity-integrated brightness temperature (in K km s$^{-1}$ units) for each pixel from the face-on view of model \texttt{lmde}, as well as Stokes $I$ and $V$ profiles taken from an example location in the envelope. 

The line shapes observed in this location are generally representative of what we see across the envelope, with roughly Gaussian Stokes $I$ emission (the particular shape of the line, however, will of course depend on the velocity structure of the chosen location, and the viewing angle).
Notably, the Stokes $V$ morphology matches well with the $dI/d\nu$ fit, which is consistent with the expectation of a fairly uniform, well-behaved line-of-sight magnetic field in the envelope (see e.g., Fig.~\ref{fig:y3dstream}).
Incidentally, the percentage polarization we obtain at this particular location is $\sim$1\%.

Using the velocity-integrated surface brightness in each pixel (i.e., the left panel of Figure~\ref{fig:lines_lmde}), we can also compare the fluxes obtained in our modeling with real observed fluxes.
Particularly, TMC-1 is a Class I source with known bright CN emission \citep{tychoniec2021}.
It has been observed to have CN $J = 2-1$ ($\nu_0 = 226$ GHz) emission with velocity-integrated surface brightness values between $\sim$50-150 mJy beam$^{-1}$ km s$^{-1}$ in its inner envelope.
Since the line transition with which TMC-1 was observed is different from that we used in our modeling (CN $J = 1 - 0$, with $\nu_0 = 113$ GHz), we convert to velocity-integrated brightness temperature for this comparison. 
Noting that the observation of TMC-1 in \citet{tychoniec2021} used a beam size of $\approx$0.16 arcsec$^2$, we calculate the velocity-integrated specific intensity and then use the Rayleigh-Jeans equation to compute the corresponding velocity-integrated brightness temperature.
The result we obtain is that 50-150 mJy beam$^{-1}$ km s$^{-1}$ corresponds to approximately 8-25 K km s$^{-1}$.
These values are roughly comparable to those found in our simulated disk-envelope system, which has velocity-integrated brightness temperatures of $\sim$8-20 K km s$^{-1}$ in the inner envelope.
The modeling we've conducted throughout this work suggests that envelope locations with CN surface brightnesses in this general range are able to yield detectable emission with fractional polarization of order $\sim$2\%.
Sources with known bright CN emission (such as TMC-1) should be prioritized when considering targets for envelope Zeeman experiments.

\section{Summary} \label{sec:sum}
In this work, we produced simulated Zeeman emission maps of the CN $J = 1 -0$ molecular line transition from MHD simulations of low-mass star formation and massive star formation regions.
We calculated the $V/I$ percentage polarization throughout the envelopes of each environment and then placed these results in context by comparing them with the local magnetic field data supplied from the 3-dimensional inputs. 
We also compared our results with nominal instrumental limits to assess the current and future feasibility of using Zeeman observations of protostellar envelopes to broadly diagnose magnetic field character during the early embedded phase of star formation. 
Our principal conclusions are summarized below:
 
\begin{enumerate}
    \item In 3D, both models contain cells with magnetic field strengths that are in principle sufficiently large to produce circularly polarized emission that may be detectable with current instruments (Table \ref{table:3dperc}).
    Intrinsically, roughly $\sim$45-60\% of cells (depending on viewing orientation) in our low mass disk envelope system and $\sim$25-45\% of cells in our massive star envelope system have local line-of-sight magnetic field strengths that exceed 3.1 mG, the line-of-sight magnetic field strength we estimate to in principle be needed to reach a polarization percentage of 1.8\%, the nominal ALMA limit.
    Furthermore, if we ignore the low density diffuse outskirts (by placing cuts at $R < 75\text{ au}$ and $ R < 2500\text{ au}$ for models \texttt{lmde} and \texttt{MSEnv}, respectively), these percentages increase to $\sim$60-70\%. 
    \item Our simulated emission maps yield pixel-by-pixel polarization results that vary significantly across the observer space.
    Each simulation contains some regions with very low polarization and others that well exceed 1.8\% (see, e.g., the final column of Figures \ref{fig:yiobs} and \ref{fig:cyobs}).
    Broadly, the low polarization regions tend to be near the edges of the simulation box or in the central, highest intensity and optical depth sight lines.
    This leaves an intermediate range (in both radius and optical depth) where high polarization cells are preferentially located.
    For our low mass disk envelope model this favorable area corresponds to the regions just beyond the optically thick disk, and for our massive protostar envelope model it mainly corresponds to the regions just surrounding the central low-polarization ``hole.''
    Though it is difficult to identify a clearly optimal optical depth that maximizes $V/I$, overall it appears that $\tau_{LC}\sim0.1-1$ tends to produce the best conditions for a protostellar envelope Zeeman experiment. 
    \item For both model types, there are significant portions of the respective envelopes that produce percentage polarizations that in principle are accessible with current instruments (i.e., above the nominal ALMA limit, see Table \ref{table:percentages}).
    For our low mass disk envelope model, we found the percentage of pixels above 1.8\% fractional polarization to be between $\sim$$20-40\%$ (depending on line-of-sight). This increases to $\sim$$40-60\%$ when the low intensity ambient material outside $R = 75\text{ au}$ is excluded.
    Meanwhile, for the massive protostellar envelope model the percentage of pixels above 1.8\% fractional polarization is $\sim$$15\%$ overall, and $\sim$$30-40\%$ when we consider only pixels within $R = 2500\text{ au}$.
    Furthermore, both models have $>90\%$ pixels above 0.18\% polarization, meaning that a factor of ten improvement in sensitivity to the magnetic field strength beyond the current ALMA limit would in principle make essentially the entirety of these envelopes accessible to Zeeman experiments. 
    One caveat that should be noted, however, is that in this work we study emission maps produced directly from radiative transfer software. This type of simulation is useful for offering theoretical guidance, but a true direct observational comparison would require producing synthetic observations [e.g., using the \textsc{CASA} observing tool \citep{casa2007}], which is outside the scope of this work.
    \item We find that the percentage polarization in a given pixel is positively correlated with the density weighted average line-of-sight magnetic field strength (first column of Figures \ref{fig:2dhist_sm} and \ref{fig:2dhist_hmw}). 
    For each of our simulations the associated mean trend line passes through a polarization percentage of 1.8\% at roughly 3.1 mG, the reference value we predicted using Eq. \ref{eq:zeemaneq}.
    \item We find that percentage polarization is positively correlated with the magnitude of the Stokes $V$ signal, especially at large $|V|$ (third column of Figures \ref{fig:2dhist_sm} and \ref{fig:2dhist_hmw}).
    This suggests that the regions that are most favorable in terms of polarization percentage are also the most favorable in terms of raw Zeeman signal.
    \item Some regions of high polarization in our massive protostar model are spatially correlated with the location of the shell we included to simulate a cavity from a protostellar wind (see first and last columns of Figure \ref{fig:cyobs}).
    Such sites are predicted to be locations of CN enhancement due to far-ultraviolet dissociation chemistry, and may therefore be good targets for Zeeman experiments with CN.
    \item Stacking the seven main sub-components of the CN $J = 1-0$ transition in our low mass disk envelope simulation yielded an average signal boost of a factor of about 8, with some pixel-to-pixel variation (between about $\sim$6-10 for the Stokes $I$ and $\sim$7-25 for the Stokes $V$).
    Stacking affects the percentage polarization results only modestly, decreasing $V/I$ in some parts of the optically thick disk and increasing $V/I$ in parts of the envelope by up to $\sim$20\% (Figure \ref{fig:stack}).
    \item Convolution of the emission from our massive protostellar envelope with a $\theta_{\rm FWHM} = 0.5^{\prime \prime}$ beam resulted in a narrower distribution of observed $V/I$ values, reducing the fraction of pixels with percentage polarizations above 1.8\% (Figure \ref{fig:conv}).
    Both the unconvolved and convolved maps have an average $V/I \sim 0.8\%$.
    These results indicate that high-resolution observations offer significant advantages to detection prospects for Zeeman experiments in protostellar envelope environments.
\end{enumerate}

\section*{Acknowledgements}
We thank the referee for a constructive report.
RRM is supported by NASA SOFIA grant 09-0117 and an ALMA SOS award.
RRM also acknowledges support from the Virginia Space Grant Consortium (VSGC) Graduate Fellowship Award, as well as helpful correspondence with Ilse Cleeves and Crystal Brogan.
ZYL is supported in part by NASA 80NSSC20K0533 and NSF AST-2307199. 
LMF acknowledges support from an NSERC Discovery Grant (RGPIN-2020-06266) as well as a Research Initiation Grant from Queen’s University.
We thank Chris McKee for helpful discussion regarding developing the \texttt{MSEnv} model.
Part of the numerical work presented in this study was supported by NASA through a NASA ATP grant NNX17AK39G (RIK) as well as the NASA High-End Computing (HEC) Program through the NASA Advanced Supercomputing (NAS) Division at Ames Research Center, and by the National Energy Research Scientific Computing Center (NERSC), a U.S. Department of Energy Office of Science User Facility located at Lawrence Berkeley National Laboratory, operated under Contract No.\,DE-AC02-05CH11231 using NERSC award NP-ERCAP0023065.
This work was performed under the auspices of the U.S. Department of Energy (DOE) by Lawrence Livermore National Laboratory under Contract DE-AC52-07NA27344 (RIK and CYC). LLNL-JRNL-855781.
The authors acknowledge Research Computing at The University of Virginia for providing computational resources and technical support that have contributed to the results reported within this publication. URL: \url{https://rc.virginia.edu}

%%%%%%%%%%%%%%%%%%%%%%%%%%%%%%%%%%%%%%%%%%%%%%%%%%
\section*{Data Availability}
The simulation data produced and analyzed for this work are available from the corresponding author, upon request.

%%%%%%%%%%%%%%%%%%%% REFERENCES %%%%%%%%%%%%%%%%%%

% The best way to enter references is to use BibTeX:

\bibliographystyle{mnras}
\bibliography{main} % if your bibtex file is called example.bib

% Alternatively you could enter them by hand, like this:
% This method is tedious and prone to error if you have lots of references
%\begin{thebibliography}{99}
%\bibitem[\protect\citeauthoryear{Author}{2012}]{Author2012}
%Author A.~N., 2013, Journal of Improbable Astronomy, 1, 1
%\bibitem[\protect\citeauthoryear{Others}{2013}]{Others2013}
%Others S., 2012, Journal of Interesting Stuff, 17, 198
%\end{thebibliography}

%%%%%%%%%%%%%%%%%%%%%%%%%%%%%%%%%%%%%%%%%%%%%%%%%%

%%%%%%%%%%%%%%%%% APPENDICES %%%%%%%%%%%%%%%%%%%%%

\appendix

%\section{} \label{sec:}

%%%%%%%%%%%%%%%%%%%%%%%%%%%%%%%%%%%%%%%%%%%%%%%%%%

% Don't change these lines
\bsp	% typesetting comment
\label{lastpage}
\end{document}